\documentclass[preprint,journal]{vgtc}       % preprint (journal style)
\usepackage{EuroVis_style/clrscode3e}
\pdfoutput=1

\usepackage{mathptmx}
\usepackage{graphicx}
\usepackage{times}

\usepackage{amsmath}
\usepackage{amssymb}
\usepackage{algorithm}
\usepackage{verbatim}
\usepackage{url}
\usepackage{multirow} 
\usepackage{xcolor}% http://ctan.org/pkg/xcolor

\usepackage{cases}
\usepackage{subfigure}
\DeclareMathOperator*{\argmax}{arg\,max}
\usepackage{EuroVis_style/clrscode3e}
\newcommand{\pluseq}{\mathrel{+}=}

\vgtccategory{Research}

\vgtcinsertpkg

\title{A Coloring Algorithm for Disambiguating Graph and Map Drawings}

\author{Yifan Hu and Lei Shi}
\authorfooter{
\item
Yifan Hu is with Yahoo Labs. E-mail: yifanhu@yahoo.com.
\item
Lei Shi is with State Key Laboratory of Computer Science, Institute of
Software, Chinese Academy of Sciences. E-mail: shil@ios.ac.cn.
}

\shortauthortitle{Hu \& Lei: A Coloring Algorithms}
\abstract{Drawings of non-planar graphs always result in edge crossings. When
there are many edges crossing at small angles, it is often difficult
to follow these edges, because of the multiple visual paths resulted
from the crossings that slow down eye movements. In this paper we
propose an algorithm that disambiguates the edges with automatic
selection of distinctive colors. Our proposed algorithm computes a
near optimal color assignment of a dual collision graph, using a novel
branch-and-bound procedure applied to a space decomposition of the
color gamut. We give examples demonstrating the effectiveness of this
approach in clarifying drawings of real world graphs and maps.
}


\begin{document}
%% ``\begin{document}'' command. It prepares and prints the title
%% block.

\firstsection{Introduction\label{sec_intro}}
\maketitle
%\section{Introduction\label{sec_intro}}

Graphs are widely used for depicting relational information among
objects. Typically, graphs are visualized as node-link diagrams \cite{Battista_etal_1999}. In
such a representation, edges are shown as straight lines, polylines or splines. Graphs that
appear in real world applications are usually non-planar. For such graphs, 
edge crossings in the layout are unavoidable.  It is a commonly accepted principle
that the number of edge crossings should be minimized whenever possible,
this principle was confirmed by user evaluations which showed that 
human performance in path-following is negatively 
correlated to the number of edge crossings \cite{Purchase_1997,Ware_2002}.
Later studies found that the effect of edge crossings varies with the crossing angle.
In particular, the task response time decreases as the crossing angle increases, and
the rate of decrease levels off when the angle is close to 90 degree \cite{Huang_crossing_2007,Huang_corssingangle_2008}. 
This implies that it is important not only to minimize the number of edge crossings, 
but also to maximize the angle of the crossings. Consequently, generating drawings that 
give large crossing angles, or even right crossing angles,
became an active area of research (e.g., \cite{Didimo_rac_2011}).
Nevertheless, for general non-planar graphs, there is no known algorithm
that can guarantee large crossing angles for straight line drawings.
Therefore, 
techniques to mitigate the adverse visual effect of small angle crossings are important in practice.

In this paper we propose to use colors to help differentiate edges. 
Our starting point is an existing  layout, and our working assumption is that
the graph is to be displayed as a static image on paper, or on screen.
The motivation comes from users of our graph drawing software.
%sfdp\cite{Hu_2005} software in Graphviz \cite{graphviz:2000}. 
These users were generally happy with the layouts of their graphs, but
were asking whether there was any visual instrument that can 
help them follow edges better.
Examining their layouts, we realized that because edges were
drawn using the same color (e.g., black), when there were a lot of
edge crossings, it was difficult to visually follow these
edges. Thus the feedback from our users, and
our own observation, echo the findings by Huang {\textit et al.}~\cite{Huang_crossing_2007,Huang_corssingangle_2008}.
When explaining why small crossing angles are detrimental to the task of following a path,
they found, with the help of  an eye tracking device, that {\em ``when edges cross
at small angles, crossings cause confusion, slowing down and triggering extra eye movements.''} and that
{\em ``in many cases, it is crossings that cause confusion, making all the paths between two nodes, and branches along
these paths, unforeseeable. Due to the geometric-path tendency, human eyes can easily slip into the edges that are close to the geometric path but not part of the target path.''}.

Edge crossing is not the only hindrance to the visual clarity of a graph drawing.
An additional problem  is that when an edge from node $u$ passes underneath the label of a node $v$ and
connects to a node $w$, it is impossible to tell visually whether there
is one edge $u\leftrightarrow w$, or two edges $u\leftrightarrow v$ and $v\leftrightarrow w$, when all edges are of
the same color (e.g., Fig.~\ref{close_edges}(b)). While these problems can be solved with user
interactions by clicking on an edge of interest, or on a node to bring
its neighbors closer (see, e.g., \cite{Moscovich_2009_linkslide}),
this involves an extra step for the user that 
may not be necessary if edges can be  differentiated with a proper
visual cue. Furthermore, there are situations
where interaction is not
possible, e.g., when looking at a static image of a graph on screen, or in print.
These are the situations that are of particular interest  in this paper.

We believe all the above mentioned problems of visually distinguishing and following edges
can be greatly alleviated by choosing appropriate colors or line styles to differentiate edges. We
first identify edge pairs that need to be differentiated (we call them
{\em colliding edges}), and
represent them as nodes of a dual collision graph. We then propose an algorithm to
assign colors to the nodes of this dual graph, in a way that maximizes the color difference
between nodes that share an edge. Thus our main contributions are:

\begin{itemize}
\item{} An approach for establishing a dual graph among colliding edges/regions,
  and coloring the nodes of the dual graph to  disambugate graph/map drawings.
\item{} A novel branch-and-bound graph coloring algorithm  that finds
  the globally optimal color embedding of each node with regard to its
  neighbors, and that works with both continuous color spaces and discrete color palettes.
\item{} A user study that establishes the effectiveness of the coloring
  approach, as well as its limitations.
\end{itemize}

We were made aware of the work of Jianu {\textit et
al.}~\cite{Jianu_2009_edge_color}, who proposed a similar idea. We
believe our work is substantially different and better than that
of \cite{Jianu_2009_edge_color}. We will discuss in more details in
the next section, and in Section~\ref{sec_results}.

%Our coloring algorithm is based
%on finding the global optimal color embedding of each node 
%with regard to its neighbors, using a branch-and-bound procedure applied to an octree space
%decomposition of the color gamut.  

%The rest of the paper is organized as follows. Section~\ref{sec_related} discusses related work.
%Section~\ref{sec_alg} defines the edge coloring problem, and CLARIFY, our algorithm for
%assigning edge colors.
%Section~\ref{sec_results} discusses color spaces and palettes, complexity, parameter choices and evaluates our algorithm experimentally.
%Section~\ref{sec_limit} discusses the limitations of our approach, and Section~\ref{sec_conc} presents a summary and topics for further study.

\section{Related Work\label{sec_related}}

Graph coloring is a classic problem in algorithmic graph theory.
Traditionally the problem is studied in a combinatorial sense. For
example, finding the smallest number of $k$ colors
on the vertices of a graph so that no two vertices sharing an edge have the same color.
%(this number is known as the chromatic number of a graph). 
 The difference between this and our work is that in
  $k-$colorability problem, a solution is valid as long as any pairs of
  vertices that share an edge have different colors, 
no consideration is given
to maximizing the actual color differences. So in
essence, the distance between colors
are binary -- either 0, or 1. For our problem we assume
that even among distinctive colors, the differences are not equal, and
are measured by color distances.
In the special case when only $k$ colors are
allowed, our algorithm
degenerate to find the optimal color assignment among all 
solutions of the $k-$colorability problem. 

This
last problem of assigning colors was also studied by Gansner~{\textit
  et al.}~\cite{Hu_2010_gmap} and by Hu~{\textit et
  al.}~\cite{Hu_2011_differential_color}, in the context of
coloring virtual maps to maximize the color difference between neighboring
regions 
In these work, maps were colored by an optimal permutation of a fixed
list of $k$
colors, with $k$ the number of countries in the map.
%Their works were motivated by the need to color virtual geographical maps, for which  a pastel color  palette is typically
%used. Thus they assumed that colors were pre-selected from a discrete palette, and arrives at the map coloring by finding an optimal permutation of the palette.
%used 
%orders induced by the maximal eigenvector of the Laplacian as the initial coloring. They
%improved the color difference further by color swapping. 
On the other hand, we assume that the color space can be either continuous
or discrete, and we select among all colors in the color space to increase color differences.

Dillencourt~{\textit et al.}~\cite{Dillencourt_Eppstein_Goodrich_color_2007} studied the problem of coloring
geometric graphs so that colors on nodes are as different as
possible. The problem they studied is very related to ours, except
that in their case the application is the coloring of geometric regions, while
we are also interested in coloring edges of a graph. 
%In addition,
%because of the nature of their application, they want {\em all} nodes to have as
%different colors as possible, while we only want nodes (of
%the dual graph) that share an edge to have as distinct colors or forms
%as possible. As a result, the objective function is different.
Dillencourt~{\textit et al.} used a force-directed gradient decent algorithm to
find a {\em locally optimal} coloring of each node with regard to its neighbors.
We 
propose a new algorithm based on a branch-and-bound process over an octree decomposition of the color space, 
that finds a {\em globally optimal} coloring for each node with regard to its neighbors.
Furthermore, our approach is more flexible and works for discrete color palettes,
in addition to continuous color spaces.

\begin{comment}
Research in choosing colors effectively for data presentation are also 
related to our work.
Healey \cite{Healey_1996_colorchoice} proposed a method for selecting multiple colors that 
provides good differentiation and allows an observer to quickly and accurately 
finding any one of the given colors.
The problem of selecting color scales and color maps for 
data visualization has also been studied by various researchers \cite{Levkowitz_1992_colorscale,Robertson_1988_color, Ware_1988_color}.
\end{comment}

Given the findings by Huang~{\textit et al.}~\cite{Huang_crossing_2007, Huang_corssingangle_2008} that 
edge crossings at close to 90  degree hamper human performance  less
than those at smaller angles, there are active researches in the so called RAC drawings of graphs.
In such a drawing, edges cross at the right angle (e.g., \cite{Didimo_rac_2011}). This is a practice employed in 
hand and algorithm drawn metro maps as well (e.g., \cite{Wolff_2007_metromap}). However, it was shown
\cite{Didimo_rac_2011} that a straight-line RAC drawing can have at most $4n-10$ edges, with $n$ the number of vertices.
As far as we are aware, even that is only a necessary, but not sufficient, condition. Therefore
techniques to help alleviate the effect of small angle crossings, when RAC or larger angle drawings are not feasible,
are important in practice.

The angular resolution of a drawing is the sharpest angle formed by
any two edges that meet at a common vertex of the drawing.  In
addition to maximizing crossing angles, for the same reason of visual
clarity, there have been researches to maximize the angular resolution
of the drawing.  Most recently, Lombardi
Drawing of graphs was proposed \cite{Duncan_2012_lombardi, Chernobelskiy_2012_lombardi}, in which edges are drawn as arcs with perfect
angular resolution. 
%This is however not possible for general
%non-planar graphs, for which a force-directed near-Lombardi drawing
%algorithm was proposed \cite{Chernobelskiy_2012_lombardi}.  
However,
Purchase {\textit et al.}~\cite{purchase_2013_lombardi} found that even though users prefer the Lombardi
style drawings, straight-line drawings created by spring-embedder
gives better performance for path following and neighbor finding tasks.  
For
straight-line drawings, while it is possible to adjust the layout to
improve the angular resolution (e.g., \cite{DiBattista_1993_angles,Garg94planardrawings}), the extent to which this can be done is
limited. 
Although previous study by Purchase~{\textit et
al.}~\cite{Purchase_2000_gdaesthetics} did not find sufficient support
for maximizing angular resolution, we do find that 
when two edges connected to the same node are almost on top to each
other, it is difficult to tell whether these are two edges or one.
%(see, e.g., Fig.~\ref{random_graph}, the two edges linking nodes 11, 7
%and 11, 8), f
For this reason we consider such edges as in
collision too.

Edge bundling is another useful tool for decluttering the tangled mess in
drawings of complex graphs \cite{Gansner_2006_improvedcircular,
Holten06_hierarchicaledge, Cui_2008_bundling, Holten_2009_FD_edgebundling,
Gansner_Hu_North_Schedigger_bundling_2011}. 
% In this approach, edges are represented by
%deformable curves, typically splines.  Edges which are in some sense
%close to each other are combined into a single bundle, sharing part of
%their routes.  Edge bundling has been proposed for circular and
%hierarchical layouts~\cite{Gansner_2006_improvedcircular,
%Holten06_hierarchicaledge} as well as for general undirected graph
%layouts~\cite{Cui_2008_bundling, Holten_2009_FD_edgebundling,
%Gansner_Hu_North_Schedigger_bundling_2011}.  
However when edges are
bundled, it is no longer possible to follow an individual edge to its
exact destination.  Pupyrev~{\textit et
al.}~\cite{Pupyrev_2012_edgebundling} proposed to separates edges
belonging to the same bundle by a small gap. While this makes it possible in theory to follow
individual edges, in practice the edges in each bundle are drawn very close to
each other. We believe whether fully bundled, or
separated by a small amount, bundled or routed 
edges can benefit from using colors to
differentiate among them (see Fig.~\ref{custa}). 
%Indeed we successfully applied our technique
%to spline edges that resulted from edge routing, a technique similar to
%edge bundling in their visual effect (see Fig.~\ref{custa}).

For directed graph drawings, a number of
studies were conducted to
evaluate the most effective way to convey the direction of edges.
Holten\ {\textit et al.}~\cite{Holten_2011_taperedges} found that
tapered edges are most effective amongst 5 different representations,
including using standard arrows. Burch~{\textit et al.}~\cite{Burch_2011_partial_edges} found 
that
partially drawn links can lead to shorter task completion times.
The idea of partially drawn edges were also used for
avoiding visual clutter at edge crossings. Rusu~\cite{Rusu_2011_gestalt} proposed a solution
of breaking edges at edge crossings to improve graph readability. 
Bruckdorfer~\cite{Bruckdorfer_2013_partiall_edges} conducted a
theoretical study of graphs that can be drawn with partial edges to avoid crossings.
Interestingly, the idea of using partially drawn edges can be dated at least as far back as 1997, when
Becker~{\textit et al.}~\cite{Becker_1995_partial_edges} visualized
the network
overload between the 110 switches of AT\&T's long distance network
during
San Francisco Bay area earthquake in 1989.
The large amount of edges occludes much of the map of the US. They solved this
problem using partially drawn edges, making the resulting picture
much clearer. 

We note that a nice way to follow an edge, or to find the neighbors of
a node, is to use interactive techniques such as ``link sliding'' and
``bring \& go'' \cite{Moscovich_2009_bring_n_go}. The algorithm we
propose is primarily aimed at disambiguating a static drawing
displayed on screen or printed on paper, it can nevertheless be used
in conjunction with such interactive techniques.

Finally, we were made aware of the work of
  Jianu {\textit et al.}~\cite{Jianu_2009_edge_color} after the completion of this work.
Jianu {\textit et al.}~\cite{Jianu_2009_edge_color} proposed a similar idea of using
  colors to differentiate edges. 
However there are multiple important differences between that work and ours.
The construction of dual
  graph is different: Jianu~{\textit et~al.} set the edge weights among all edges to be
the inverse of  either the intersection angle, or the edge distance if the edges do
  not intersect, which is not optimal since it is perfectly harmless to
  color edges that have no conflict
  with the same color. In fact,
  their method always results in a complete dual graph, making it more expensive
  for relatively large graphs. Furthermore, because of the
  complete dual graph, all edges of the original graph must have
  different colors. Therefore the
  drawings in~\cite{Jianu_2009_edge_color}, which are all of very
  small graphs, always contains a multitude of colors, which is unnecessary.
  Our collision
  graph almost always contains disconnected components (e.g.,  Fig.~\ref{dual}).
  This decomposes the coloring problem into smaller ones, and
  allows us to use the same (black)
  colors for many edges.
Jianu et al.~\cite{Jianu_2009_edge_color} solved the coloring problem
using a force-directed algorithm, motivated by Dillencourt~{\textit
  et al.}~\cite{Dillencourt_Eppstein_Goodrich_color_2007}. 
We obtained the code for \cite{Jianu_2009_edge_color} from one of the authors.
Based on reading the code, we found that it applies force directed
algorithm to nodes of the dual graph in the 2D subspace of the LAB
color space (the AB subspace). It then sets a fixed L value of 75 (L
is the lightness, between 0 to 100). This observation is
consistent with the drawings in~\cite{Jianu_2009_edge_color}, where
black background is used for all drawings due to the high lightness
value
(see also Fig.~\ref{palettes}(d)).
This makes the algorithm limited to a small subset of all possible
colors.
Finally, the force-directed algorithms of Dillencourt~{\textit et
  al.}~\cite{Dillencourt_Eppstein_Goodrich_color_2007} and 
Jianu~et~al.~\cite{Jianu_2009_edge_color}
can only be applied to continuous color space in 2D or 3D. Neither works for user specified 
color palettes, or 1D colors. Our algorithm works for both
continuous or discrete color spaces.
Overall, we believe that the idea of using colors for
disambiguating edges are quite natural to think of. 
It is how to use the appropriate algorithm to make the idea work
effectively in practice that is crucial and that differentiates our work and \cite{Jianu_2009_edge_color}.
Furthermore,  we present a first user study to evaluate this idea with
real users. The results suggest possible scenarios when the edge
coloring approach is effective.

\begin{figure}[h]
\vspace{-.0cm}
\begin{center}
\hbox{\hspace{-.0cm}\includegraphics[width=0.53\linewidth]{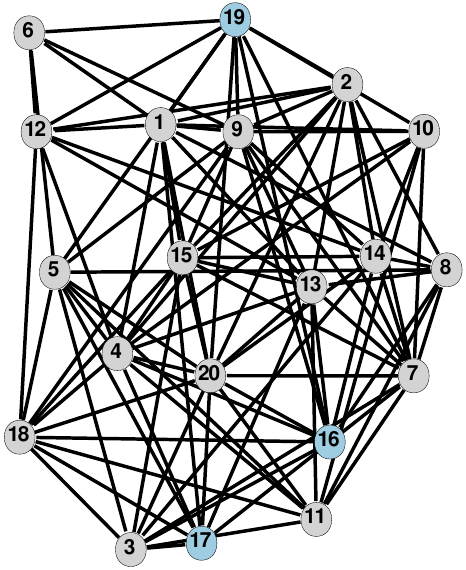}\hspace{-0.25cm}\includegraphics[width=0.53\linewidth]{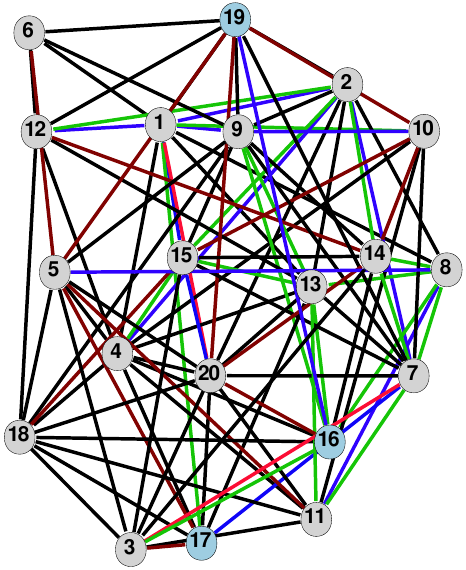}}
\end{center}
\vspace{-1cm}
\caption{\textsf 
Left: a graph with 20 nodes and 100 edges. It is difficult to follow
some of the edges. For example, is node~19 (blue) connected to node~16 (blue)? Is node 19 connected to 17 (blue)? Right: the same graph, with the edges colored using our algorithm. Now it is easier to see that 19 and 16 are connected by a blue edge, but 19 and 17 are not connected.
\label{random_graph}}
\vspace{-.5cm}
\end{figure}

\section{The Edge Coloring Problem and a Coloring Algorithm\label{sec_alg}}

Appropriate coloring can help greatly in differentiating edges that
cross at a small angle.
Fig.~\ref{random_graph}~(left) illustrates such a situation.  With
many crossing edges, it is difficult to follow the edge from node 19 (top-middle, blue)
to node 16 (lower-right, blue).  In comparison, in Fig.~\ref{random_graph}~(right), it is
easier to see that 19 is connected to 16 by a blue edge. The objective of this section is
to identify situations where ambiguities in following edges can occur, and propose an edge coloring algorithm 
 to resolve such ambiguities.

\definecolor{lightgray}{rgb}{1,1,0.96}
\definecolor{lightgray2}{gray}{1}
\setlength\fboxsep{-3pt}
\begin{figure*}[t]
\def\tabcolsep{1pt}
\begin{center}
\begin{tabular}{cccc}
\hspace{-.2cm}\fcolorbox{lightgray2}{lightgray}{\includegraphics[width=8cm]{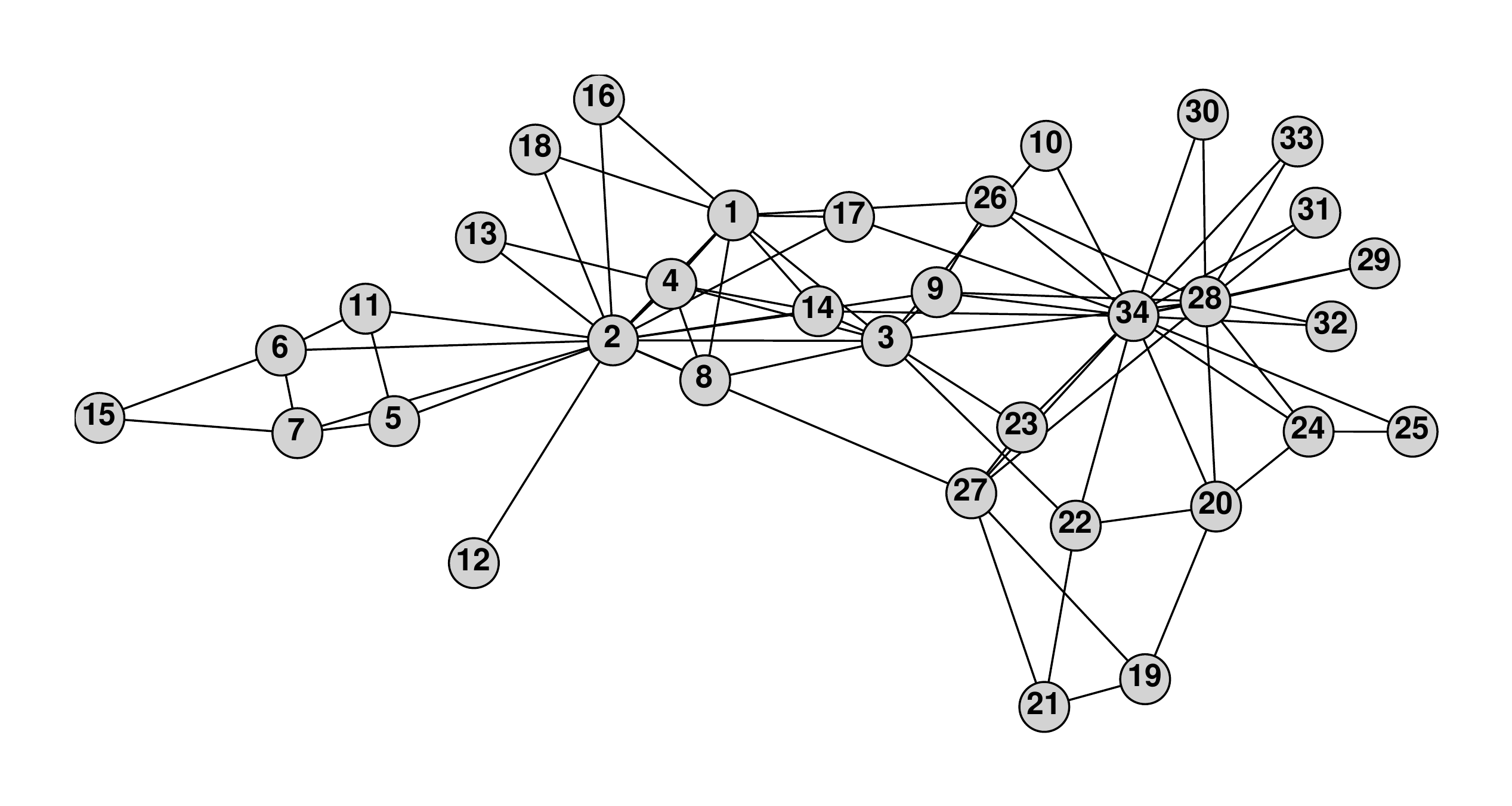}    }  
&\hspace{-.1cm}\raisebox{2cm}{$\rightarrow$} 
&\hspace{-.1cm}\fcolorbox{lightgray2}{lightgray}{\includegraphics[width=8cm]{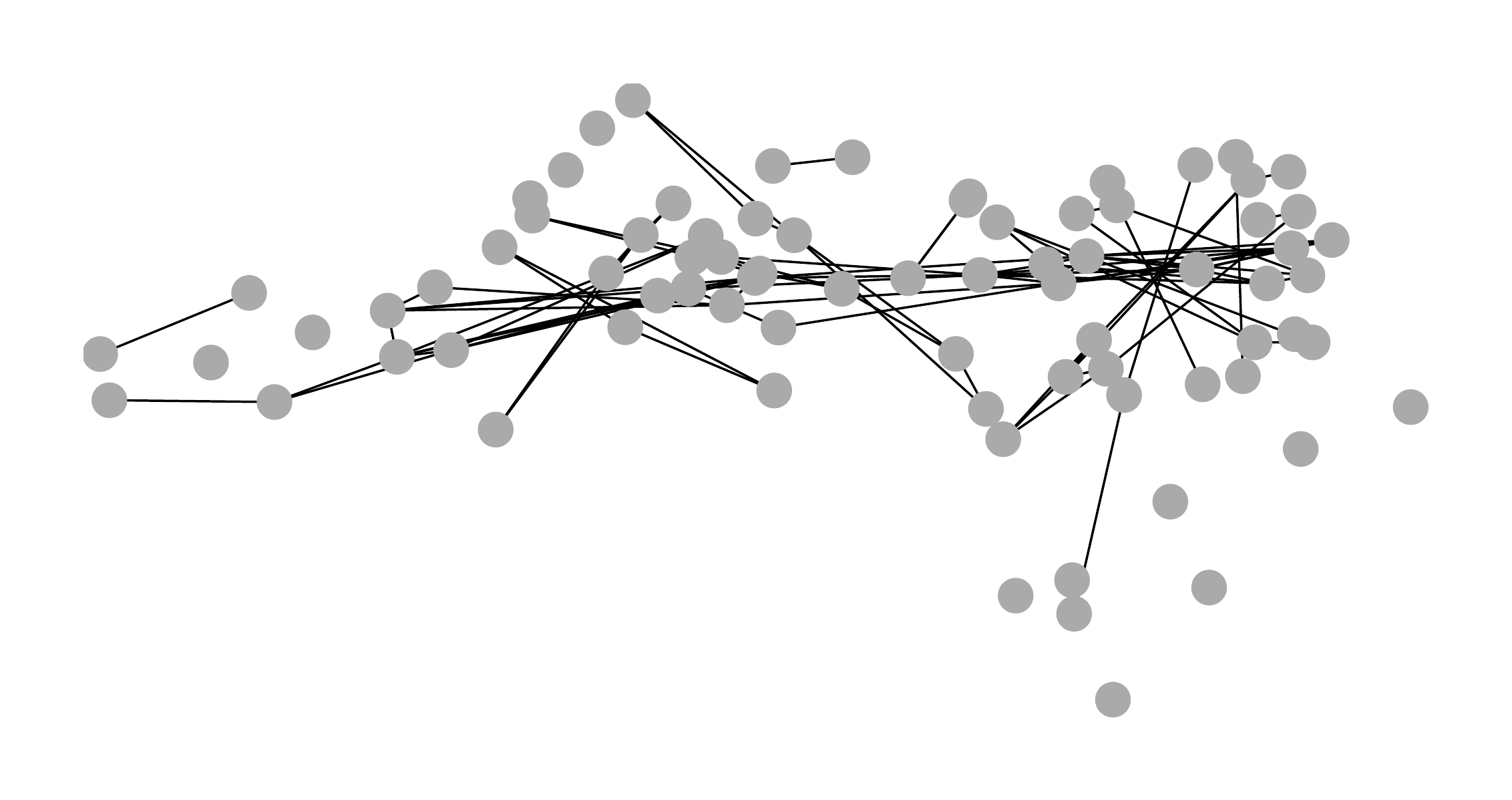}}
&\hspace{-.1cm}\raisebox{2cm}{$\rightarrow$}\\
\hspace{-.2cm}(a) original graph& &(b) dual collision graph&\\
\hspace{-.1cm}\fcolorbox{lightgray2}{lightgray}{\includegraphics[width=8cm]{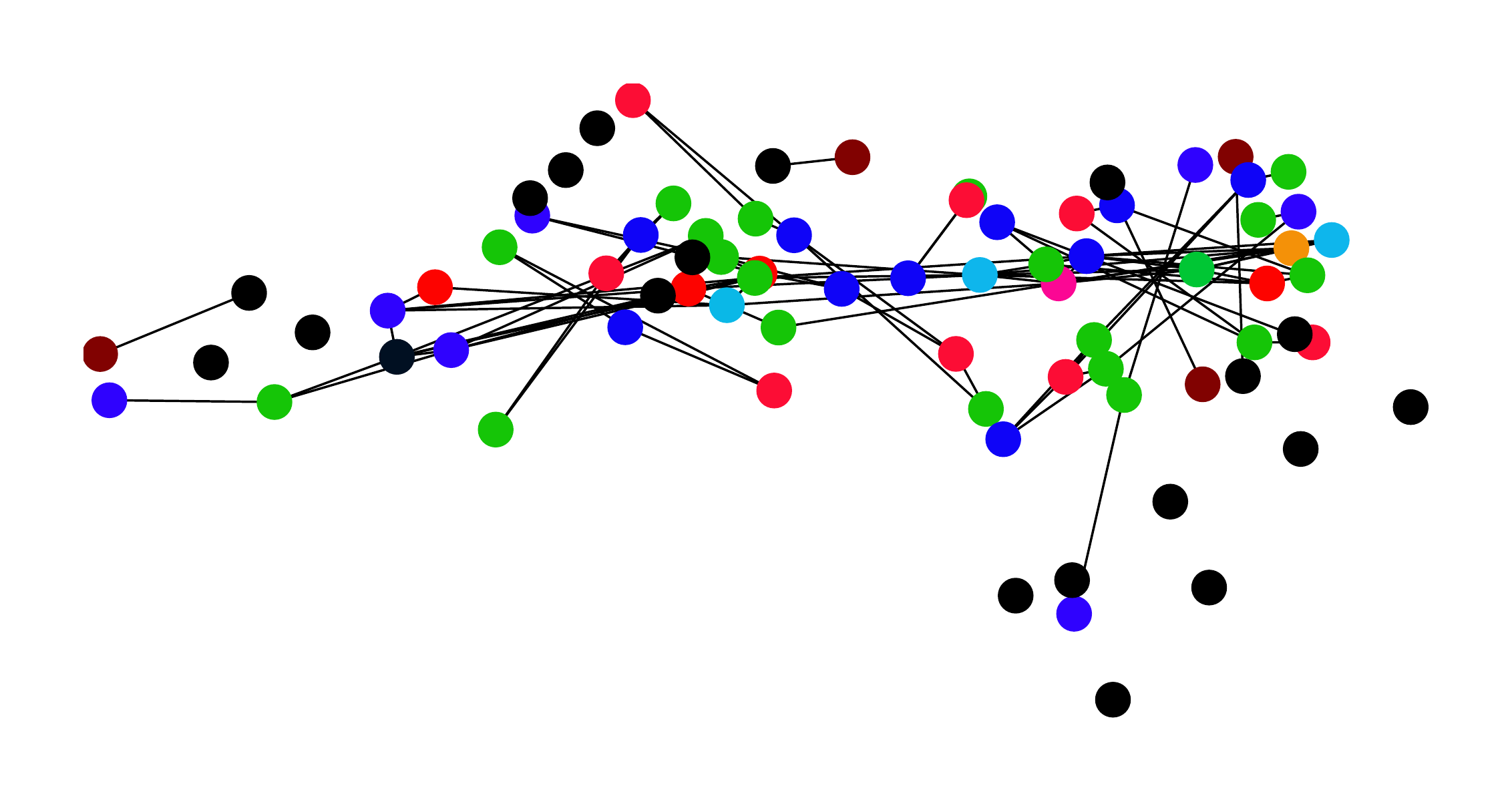} }
&\hspace{-.1cm}\raisebox{2cm}{$\rightarrow$} 
&\hspace{-.1cm}\fcolorbox{lightgray2}{lightgray}{\includegraphics[width=8cm]{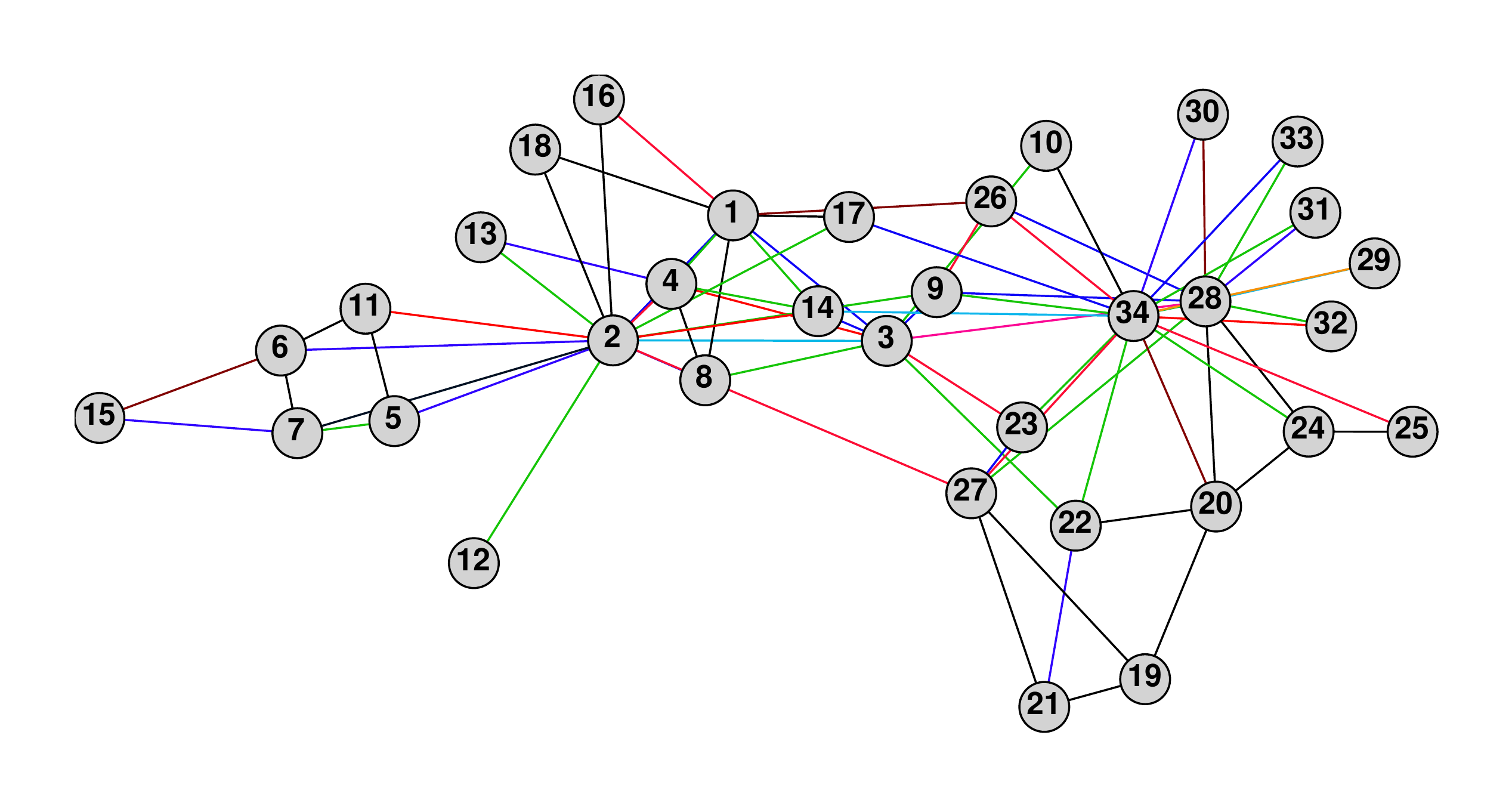}}\\
(c) dual graph colored& &(d) original graph colored\\
\end{tabular}
\end{center}
\vspace{-.5cm}
\caption{\textsf The proposed pipeline for coloring the edges of the Zachary's Karate Club Graph: (a) the original graph; (b) the dual collision graph, with each node representing an edge of the original graph, and positioned at the center of that edge; (c) the dual graph, with nodes colored to maximize color differences along the edges (see Fig.~\ref{dual} for a clearer force-directed layout of this graph);
(d) the original graph, with edges colored using the node coloring in (c).\label{pipeline}}
\vspace{-.2cm}
\end{figure*}

\subsection{Edge collisions}

Two edges are considered in collision if an ambiguity arises when they are drawn using the same color. 
The following are four conditions for edge collision:

\begin{itemize}
\item{} \emph{C1: they cross at a small angle.}

\item{} \emph{C2: they are connected to the same node at a small angle.}

\item{} \emph{C3 (optional): they are connected to the same node at an angle close to 180 degree.}

\item{} \emph{C4: they do not cross or share a node, but are very close to each other and are almost parallel.}
\end{itemize}

We now explain the rationale for considering each of these four conditions as being in collision. 
C1 is considered a collision following the user
studies described in Section~\ref{sec_intro} by Huang {\textit et
al.}~\cite{Huang_crossing_2007, Huang_corssingangle_2008}. When eyes 
try to follow an edge to its destination, small 
crossing angles between this edge and other edges create multiple paths along the
direction of the eye movement, either taking eyes to the wrong path,
or slowing down the eye movement. C2 creates a situation where one edge is almost on top
of the other, making it difficult to visually follow one of these
edges. 

C3 could create confusion as to whether the two edges connected at close to
180 degree are one edge, or two edges, when node labels are drawn. For example in
Fig.~\ref{random_graph}~(left), it is difficult to tell whether nodes
19 and 17 are connected, or whether 19 is connected to 20 and 20 is
connected to 17. When edges are properly colored (Fig.~\ref{random_graph}~(right)), it 
is clear that the latter is true. Note that if edges are allowed to be drawn on top of nodes,
then an edge between 19 and 17 would be seen over the label of 20, thus
this kind of confusion can be eliminated. Therefore we
consider C3 as optional. But drawing edges
over the label of nodes does introduce extra clutter and make it harder to read the node labels.% (see Fig.~\ref{yeast}).

\begin{figure}[htbp]
\begin{center}
\begin{tabular}{ccc}
\hspace{-.3cm}\includegraphics[width=0.24\textwidth]{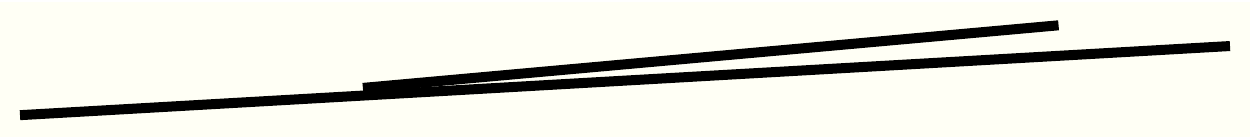}&\hspace{-.3cm}\includegraphics[width=.24\textwidth]{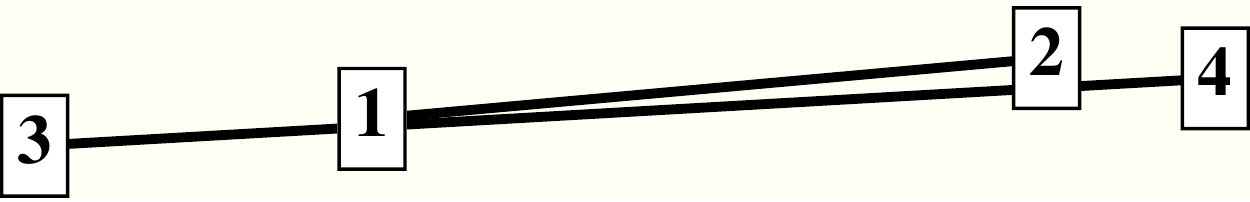}\\
(a)&(b)\\
\hspace{-.3cm}\includegraphics[width=.24\textwidth]{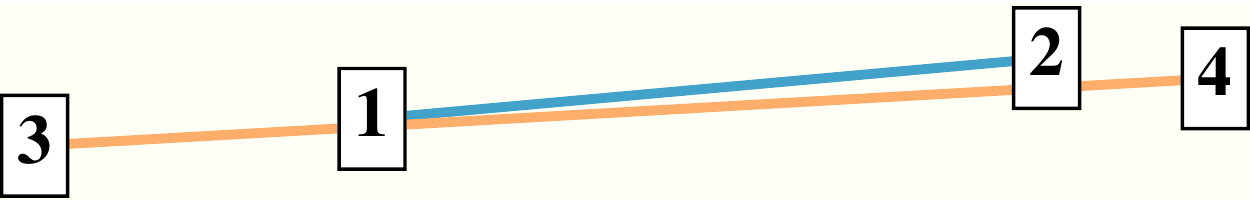}&\\
[-.5cm]\\
(c)&\\
\end{tabular}
\end{center}
\caption{\textsf An illustration of the rationale for collision condition C4. (a) Two edges that do not cross. (b) When nodes are shown, it is difficult to tell if there are two edges ($1\leftrightarrow 2$ and $3\leftrightarrow 4$), or three edges ($1\leftrightarrow 2$, $1\leftrightarrow 4$ and $1\leftrightarrow 3$), or whether there even exists an edge $3\leftrightarrow 2$. (c) After coloring each edges with a distinctive color, it is clear that there are two edges, $1\leftrightarrow 2$ and $3\leftrightarrow 4$ \label{close_edges}}
\end{figure}

C4 causes a problem because when two edges are very close and almost
parallel, it is difficult to differentiate between them. In
addition, it can cause confusion when node labels are drawn. 
Fig.~\ref{close_edges}(a) shows two lines very close and
almost parallel. While it is possible to differentiate between the
two edges, when node labels are added (Fig.~\ref{close_edges}(b)), it
is difficult to tell whether there are two edges ($1\leftrightarrow 2$
and $3\leftrightarrow 4$), or three edges ($1\leftrightarrow 2$,
$1\leftrightarrow 4$ and $1\leftrightarrow 3$), or whether there 
even exists an edge $3\leftrightarrow 2$. 
 This confusion can be avoided if suitable edge coloring is applied (Fig.~\ref{close_edges}(c)).

To resolve these collisions, we propose to color the edges so that 
any two edges in collision have as different colors as possible.
We first construct a dual edge collision graph.

\subsection{Constructing the dual collision graph}

Let the original graph be $G=\{V, E\}$.
Denote by $N(v)$ the set of neighbors of a node $v$.
The {\em dual collision graph} is $G_c =\{V_c, E_c\}$, where each node in $V_c$
corresponds to an edge in the original graph. In other word, there is a one-to-one mapping $e: V_c\rightarrow E$.
Two nodes of the dual graph $i$ and $j$ are connected if $e(i)$ and $e(j)$ collide in the original graph.

The problem of coloring the edges of $G$ then becomes that of coloring nodes of the dual graph $G_c$. 
Let $\mathcal C$ be the color space, and  $c(i)\in \mathcal C$ be the color
of a node $i\in V_c$, we want to find a coloring scheme such that the color of each node in the dual graph
is as different to its neighbors as possible. This task can be posed as a MaxMin optimization problem:

\vspace{-.3cm}
\begin{equation}
\argmax_{c: V_c\rightarrow {\mathcal C}} \min_{\{i,j\}\in E_c} w_{ij} \|c(i) - c(j)\|,\label{optimization}
\end{equation}
\vspace{-.3cm}

\noindent  where $w_{ij}>0$ is a weight inversely proportional to how important it is to
differentiate colors of nodes $i$ and $j$, and
$\|c(i) - c(j)\|$ is a measure of the difference between the colors assigned to the two nodes.

Note that (\ref{optimization}) is stated rather generally: $\mathcal C$ could be a discrete, or continuous, color space.
This is intentional since we are interested in both scenarios. %But for the remainder of this section,
%we shall assume $\mathcal C$ is a connected region in a Euclidean
%space of dimension~$d$.
%we shall assume $\mathcal C$ is a connected region in a Euclidean
%space of dimension~$d$.
All we assume is that $\mathcal C$ sits in a Euclidean space of dimension~$d$.

\begin{figure}[htbp]
\vspace{-.5cm}
\begin{center}
\hspace{-.8cm}\includegraphics[width=1.1\linewidth]{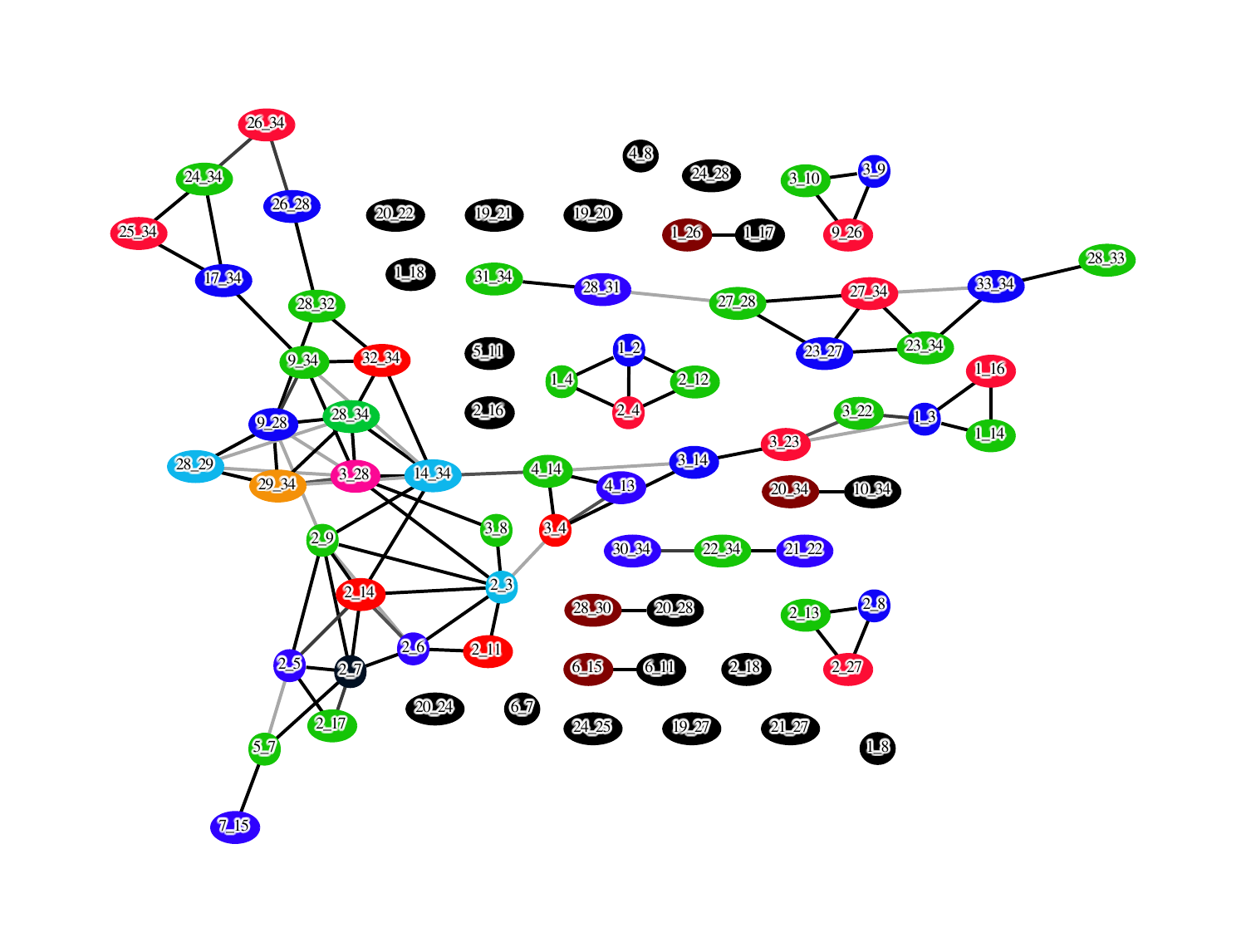}
\vspace{-1.2cm}
\caption{\textsf The dual graph in Fig.~\ref{pipeline}(c), with a force-directed layout. 
A node labeled ``$i\_j$'' represents
edge $i\leftrightarrow j$ in the original graph. Nodes are colored using Algorithm~1,
so that each node is colored as differently from its neighbors as possible.
To disambiguate edges we color them in gray scale. \label{dual} }
\end{center}
\end{figure}

Once we colored the dual graph, we can use the same coloring scheme for the edges of the original graph.
The complete pipeline of our proposed approach is  illustrated in Fig.~\ref{pipeline}. Notice that 
the dual graph in Fig.~\ref{pipeline}(b) (displayed more clearly using a force-directed layout in Fig.~\ref{dual}) is disconnected. 
We apply our algorithm on each component of the dual graph. 
%For a singleton component
%we assign a black color, and for a component of 2 nodes we assign black and dark red colors.

\subsection{A color optimization algorithm\label{color_opt}}

Dillencourt~{\textit et al.}~\cite{Dillencourt_Eppstein_Goodrich_color_2007} proposed a force-directed algorithm in a Euclidean color
space. They wanted {\em all} pairs of nodes to have
distinctively
different colors. %therefore  the dual graph in their case is  a complete graph.
Consequently their  algorithm used a force model where repulsive forces exist among all pairs of nodes.
\begin{comment}
Starting from a random embedding in the color space, at each iteration it attempts to change the position of the
vertices, one  at a time, by three types of moves: random
jumps, swaps with other vertices, and moving in the gradient
direction. For each of these three types of moves their algorithm accepts the move
only when it improves the overall quality of the coloring. If an iteration fails
to produce any quality improvement, the algorithm reduces the step size. It terminates 
when this step size falls below a preset threshold. If a move
takes the point outside of the valid color space, the algorithm pulls it back to the boundary of the space.
\end{comment}

Because in our case edges can have the same color as long as they do not collide, there is no need
to push all pairs of nodes of the dual graph apart in the color space. Therefore we can not 
use the algorithm of Dillencourt~{\textit et al.}~\cite{Dillencourt_Eppstein_Goodrich_color_2007} as is.
Although it is possible to adapt their algorithm, we opt to propose an alternative algorithm.
One reason is that we like to be able to use not only continuous color spaces, 
but also discrete color palettes (Sec.~\ref{sec_palette}).
Another reason is due to the fact that even when deciding the optimal color for one node of the dual graph with regard to 
all its neighbors, this seemingly simple problem can have many local maxima.

\begin{figure}[htbp]
\begin{center}
\hbox{\includegraphics[height=0.473\linewidth]{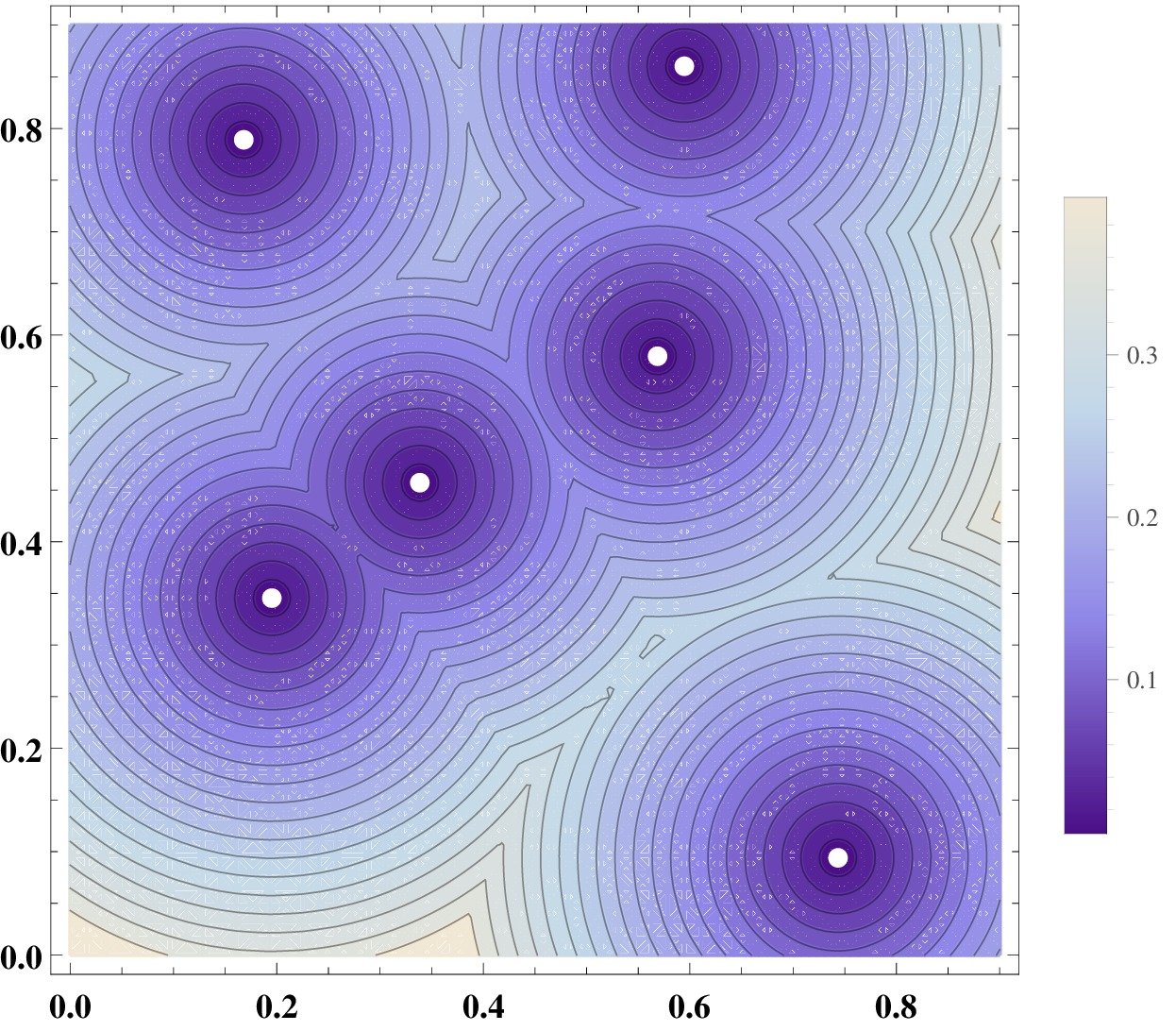}\includegraphics[height=0.47\linewidth]{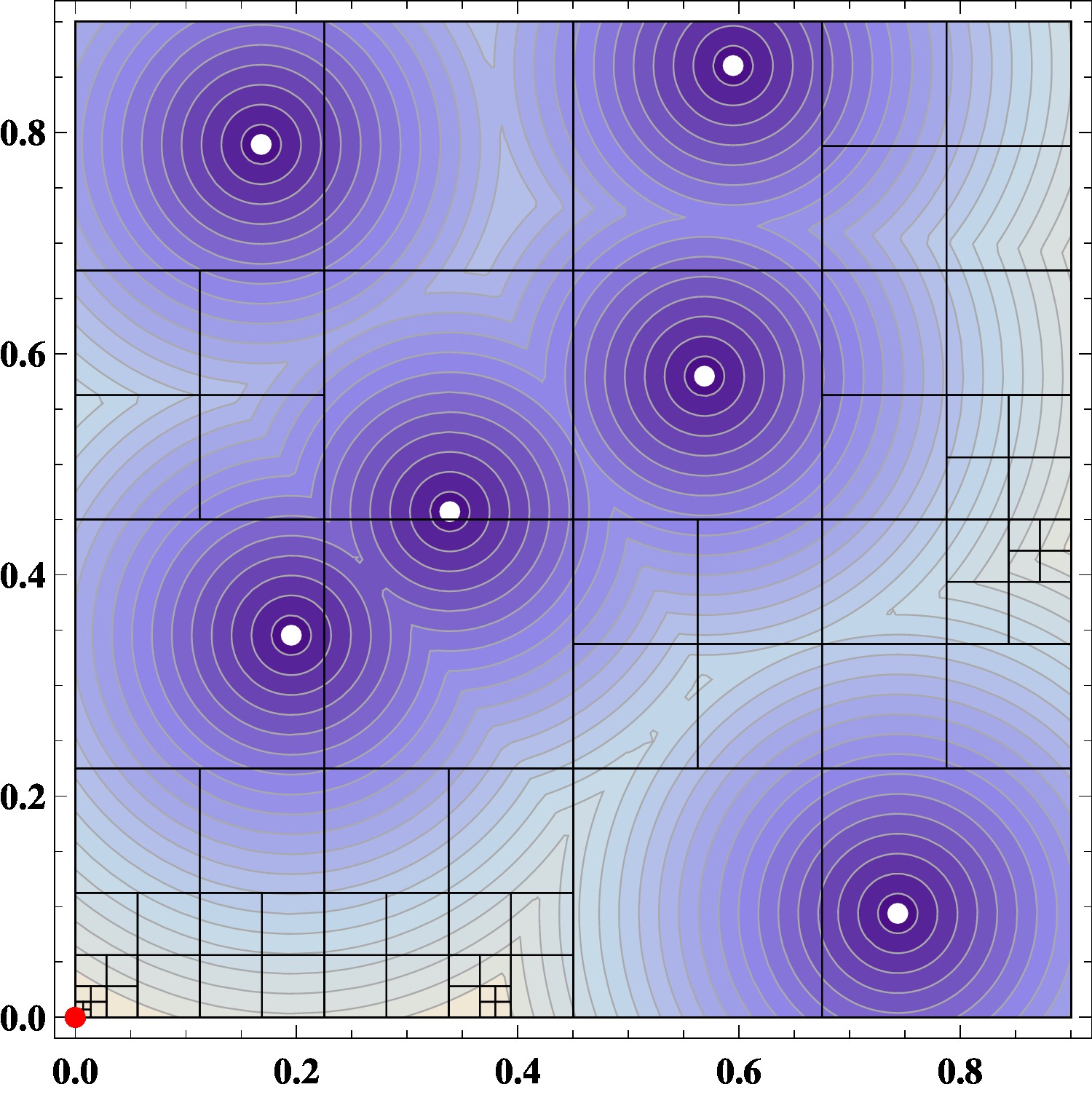}}
\end{center}
\caption{\textsf Left: contour plot of the distance to a set of six (white) points in the space $[0,0.9]\times[0,0.9]$. There are seven or more local maxima. E.g., near $\{0,0.55\}$, $\{0.35,0.9\}$ and $\{0.4,0.7\}$. Right: an illustration of the quadtree structure generated during our algorithm for finding the global optimal embedding of a node that is farthest away from the set of six points. The final solution is $\{0,0\}$ (shown as the  red point). \label{contour}}
\end{figure}

As an example, for simplicity of illustration 
we assume that our color space is 2D, and that the color distance is the Euclidean distance. Suppose
we want to find the best color embedding for a node $u$ in the dual graph with six neighbors, and the six neighbors
are currently embedded as shown in Fig.~\ref{contour}~(left). We want to place $u$ as far away from 
the set of six points as possible. Fig.~\ref{contour}~(left) shows a color
contour of the distance from 
the set of six points (the distance of a point to a set of point is defined as the minimum distance between this point and
all the points in the set, assuming unit weighting factors). Color scale is given in the figure, with blue for low values and off-white for large. From the contour plot it is clear that there are seven or more 
local maxima. In 3D there could be even more local maxima.
A force-directed algorithm such as \cite{Dillencourt_Eppstein_Goodrich_color_2007}, even with the random jumps and swaps,
is likely to settle in one of the local maxima.

Instead we hope to find the global maximum. A naive way to find the global
maximum position in the color space with regard to a set of points
is to search exhaustively by imposing a fine grid over the color space, 
and calculating the distance from each mesh point to the set. However,
given that the color space are typically of three dimensions, even at a resolution of 100 subdivisions along
each dimension, we need $10^6$ distance calculations. This is computationally too expensive, bear in mind
that this computation needs to be performed for each and every node
of the dual graph repeatedly until the overall embedding in the color space converges.

We propose a more efficient algorithm based on the octree data structure (quadtree
for 2D) that does not require evaluations of the distance over all mesh points. 
Using Fig.~\ref{contour}~(left) as an example, we want to 
find a point in the color space that is of maximal distance to a target set of points.
Define the objective function value of a square to be the distance from the center of the square to the target set.
We start with a queue of one square
covering the color space, and define the current optimal value as the maximal distance over
all squares in the queue to the target set.
Taking a square from the current queue, we subdivide it into four squares. 
If the distance of one of the four square to the point set, plus
the distance from the center of the square to a corner of the square, is
less than the current optimal distance, this square is discarded. This is because no point in this square can have
a larger distance to the target set than the current optimal distance. If the 
square is outside of the color space, it is also discarded.
Otherwise the square is entered into the queue, and the optimal value updated. 
This continues until the half width of all squares in the queue are smaller than a preset threshold $\epsilon$.
The point that achieves the current optimal value is taken as the optimum. We know that
the current optimal value should be within a value $\delta= d^{1/2} \epsilon$ to the global optimal value,
where $\delta$ is the half diagonal of the final square in $d$-dimensional space.

\begin{algorithm}[H]
{
  \caption{CLARIFY$(G, \mathcal C, \epsilon)$}
\label{alg}
\begin{codebox}
%\Procname{CLARIFY$(G, \mathcal C, \epsilon)$}
\li input: graph $G=\{V, E\}$, color space $\mathcal C$, threshold $\epsilon$
\li compute a dual collision graph $G_c=\{E_c, V_c\}$ of $G$
\li randomly choose $c(i)$ in $\mathcal C$ for all $i\in V_c$
\li set: $\text{mindist} = 0$, $\text{sumdist} = 0$
\li \Repeat
\li  set: $\text{mindist}_\text{old} \gets \text{mindist}$, $\text{sumdist}_\text{old} = \text{sumdist}$\\ \hspace{1cm} $\text{mindist} = \infty$, $\text{sumdist} = 0$
\li  \For $i \in V_c$ \Do
       \li  define $c(N(i)) := \{c(j)|j\in N(i)\}$
       \li $c(i) = \text{EmbedOneNode}(c(N(i)), \epsilon)$
       \li $\text{mindist} = \min\left\{\text{mindist},\ dist(c(i), c(N(i)))\right\}$
       \li $\text{sumdist} \pluseq dist(c(i), c(N(i)))$
     \End
\li \Until ($\text{mindist} < \text{mindist}_\text{old}$  $||$ \\ \hspace{.5cm} ($\text{mindist} = \text{mindist}_\text{old}$ \&\& $\text{sumdist} \le \text{sumdist}_\text{old}$))
\li return: $c(e(i)) = c(i),\ i\in V_c$
\end{codebox}
}

\end{algorithm}

\vspace{-.7cm}
\begin{algorithm}[H]
{
  \caption{EmbedOneNode$(C, \epsilon)$\label{alg2}}
\begin{codebox}
%\Procname{CLARIFY$(G, \mathcal C, \epsilon)$}
\li input: a set of points $C\in \mathcal C$, a  threshold $\epsilon$
\li set: $s$ a square/cube covering the color space $\mathcal C$
\li set: a first-in-first-out queue $Q=\{s\}$
\li set: $c^* = \text{center}(s)$
\li set: $dist^* = \text{dist}(s, C)$
\li \For $s\in Q$\Do
       \li \If $w(s) < \epsilon$ break
       \li $Q := Q -\{s\}$
       \li \For $t\in \text{children}(s)$\Do
       \li  \If $t\cap \mathcal C = \emptyset$ $|$ $\text{dist}(t, C) + d^{1/2} w(t) < \text{dist}^*$ \Then
               \li continue
         \End
         \li \If{ $\text{dist}(t, C) > \text{dist}^*$} \Then
             \li $c^* = \text{center}(t)$
             \li $\text{dist}^* = \text{dist}(t, C)$
         \End
	 \li $Q:=Q\cup \{t\}$
       \End
    \End
    \li return: $c^*$
\end{codebox}
}
\end{algorithm}

This algorithm is in essence a branch-and-bound algorithm operating on 
the octree (quadtree for 2D) decomposition of the color space.
When applied to the problem in Fig.~\ref{contour}~(right), we can see that in the top-left
quadrant, the quadtree branched twice and stopped, because the function values are relatively small
in that quadrant. The top-right and bottom-right quadrants branched 3 and 4 times, respectively.
The final optimal point is found in the bottom-left quadrant. Initially
the algorithm homed in on two regions, one around $\{0.375, 0\}$ and
the other around $\{0,0\}$, eventually settled around the latter.

Of course this branch-and-bound algorithm only finds the global optimum embedding
for one node. After applying the algorithm to every node of the dual graph once (one outer iteration),
if the minimal
color difference increases,
or if it does not change, but the 
total sum of color difference across all nodes increases, we
repeat. 

 We name the
algorithm CLARIFY (Edge Coloring for {\em CLARIFY}ing a Graph Layout)
and formally state it in Algorithm~1. %\ref{alg}.
First, we give some notations used in the presentation of the algorithm.
For a point $x$ and a finite point set $C$ in the Euclidean color space $\mathcal C$, we define the
 point-set distance as $\text{dist}(x, C) = \min_{y\in C} w_{x,y} \|x-y\|_2$.
We denote the center of a square or cube $s$ as $\text{center}(s)$, its children (by dividing 
a square into 4  or a cube into 8) as $\text{children}(s)$, and its {\em half width} as $w(s)$. We define the distance
between $s$ and a set of point $C$ as that between the center of $s$ and $C$, that is, $\text{dist}(s, C) = \text{dist}(\text{center}(s), C)$.
%\begin{equation*}
%\text{dist}(s, C) = \text{dist}(\text{center}(s), C).
%\end{equation*}

\noindent The CLARIFY algorithm utilizes the global optimization algorithm for embedding one node, given in Algorithm~2 %\ref{alg2}
as  $\text{EmbedOneNode}$.

\section{Implementation and Results\label{sec_results}}

We now give details on the implementation of CLARIFY, and results of using
the algorithm on real world graphs.

\subsection{Color space\label{sec_palette}}

CLARIFY works for both continuous color spaces (as long as it is a metric space),
as well as discrete ones.

\textbf{The RGB color space.}
An often used color model is RGB. This model defines color
by a combination of three color intensities, red, green, and blue.
Thus colors in the RGB model can be considered as residing in a
three-dimensional cube.
\begin{comment}The choice of these three primary colors is
related to the fact that human eyes normally have three kinds of cones
(a type of photo-receptor cells). The L cones respond the most to long
wavelength, peaking at a reddish color of around $575 \mu m$. The M
cones respond the most to medium wavelength, peaking at a green color
of around $545 \mu m$, and the S cones respond the most to short wave
length, peaking at a bluish color of around $450 \mu m$ \cite{Stockman_1993_vision}.
Colors of other wavelength will cause responses from the three types
of cones, allowing the brain to differentiate among different
colors. This also means that colors outside the response ranges of the
cones will be invisible to human eyes.

 If we consider
the intensity of each color as between 0 to 1, then the RGB color
space is the unit cube in 3D.
Assuming that a white background is used to draw the graph,
then we do not want edges to be very bright. Hence when using the RGB space,
by default we restrict the algorithm to
with $0 \le r,g,b \le 0.7$, assuming that 1 is the fully saturated color value.
\end{comment}

\textbf{The LAB color space.} RGB color model is widely used for the representing and displaying
images in electronic systems, such as LCD/LED display. 
However, distance between two colors in the RGB space is not an accurate measure of
perceived difference by human eyes. For that purpose, 
the LAB color
model is consider better \cite{lab_color2}.

\begin{comment}
The LAB color model is a three axis system that is device independent. The
 L axis stands for lightness and is between 0 to 100, the A axis 
goes from green to  magenta/red 
and the B axis goes from yellow to blue, both are between -128 to 128.
The LAB color space  is perceptually more uniform than RGB space, in the sense that
a change of the same amount in a color value should produce a change of 
about the same visual importance. 
The LAB color space includes all perceivable colors, and more.
When mapping RGB to LAB, the resulting gamut, hereafter referred to 
simply as the LAB color gamut, has a complex polygon like shape. %~(Fig.~\ref{lab}).
Mapping from RGB to LAB is unique, while that from LAB to RGB is not, due to
the fact that the number of colors representable by the LAB model exceeds that of the RGB color model,
and for a LAB color that is not representable in RGB, typically the closest RGB color is used for the mapping.
\end{comment}

The LAB color space  (a rectangular box $[0,100]\times[-128,128]\times[-128,128]$) includes all perceivable colors, and more.
While we can use CLARIFY directly in the LAB color space, since eventually we need to 
render the resulting drawing on screen and in print, we need to convert the 
coloring to RGB. Therefore we need to work within the LAB color gamut -- the part of the LAB space that corresponds to the RGB space. It has a complex shape.
Applying CLARIFY requires checking whether a cube is outside of the 
LAB gamut, which is considerably more complicated than checking whether a point is outside of the gamut.
%since we are not aware of an analytical closed-form expression of the LAB gamut. 

\begin{figure*}[t]
\vspace{-.4cm}
\def\tabcolsep{1pt}
\begin{center}
\begin{tabular}{ccc}
\hspace{-.5cm}\includegraphics[width=.35\linewidth]{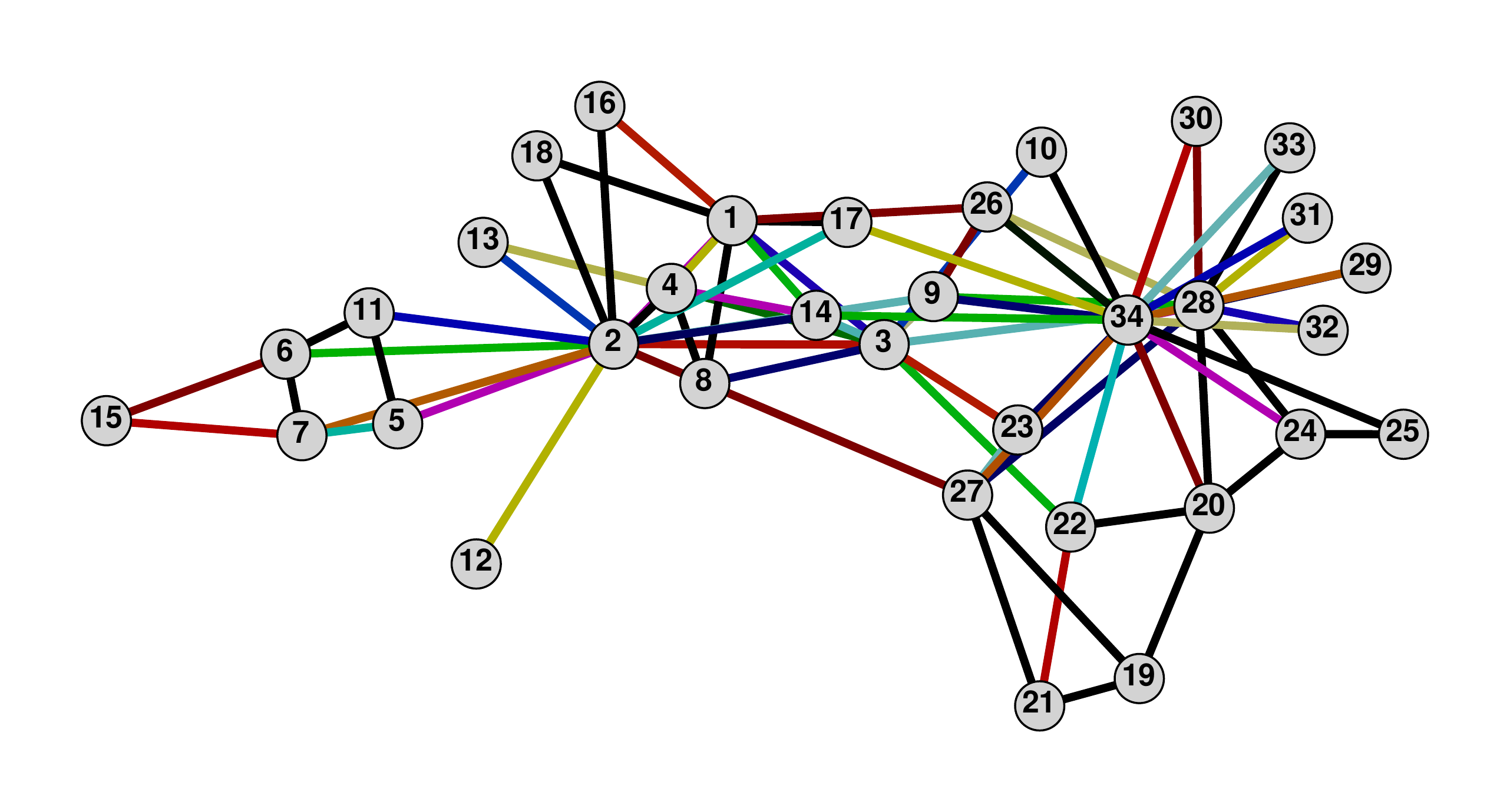}
&\hspace{-.8cm} \includegraphics[width=.35\linewidth]{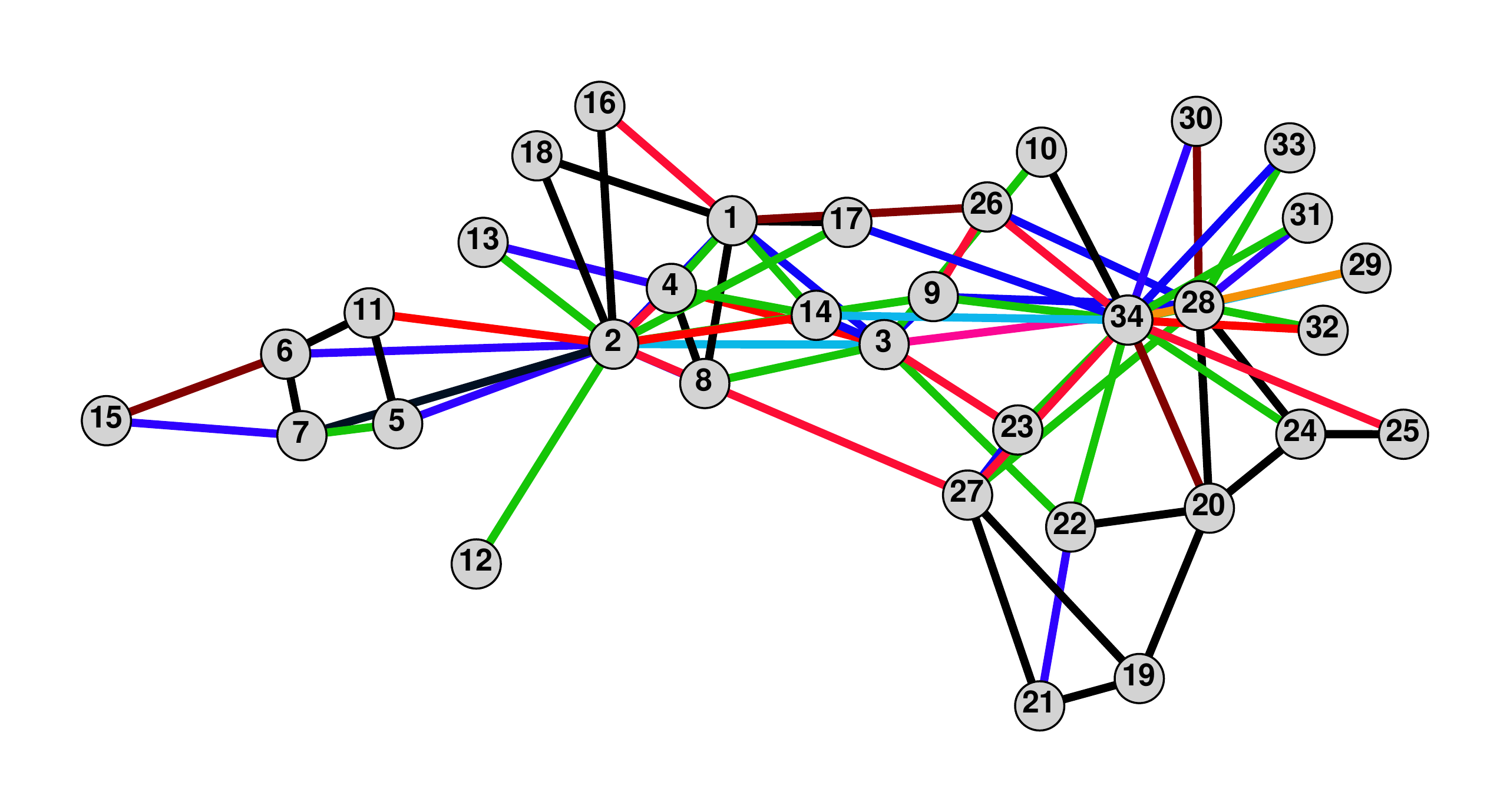}
&\hspace{-.7cm}\includegraphics[width=.35\linewidth]{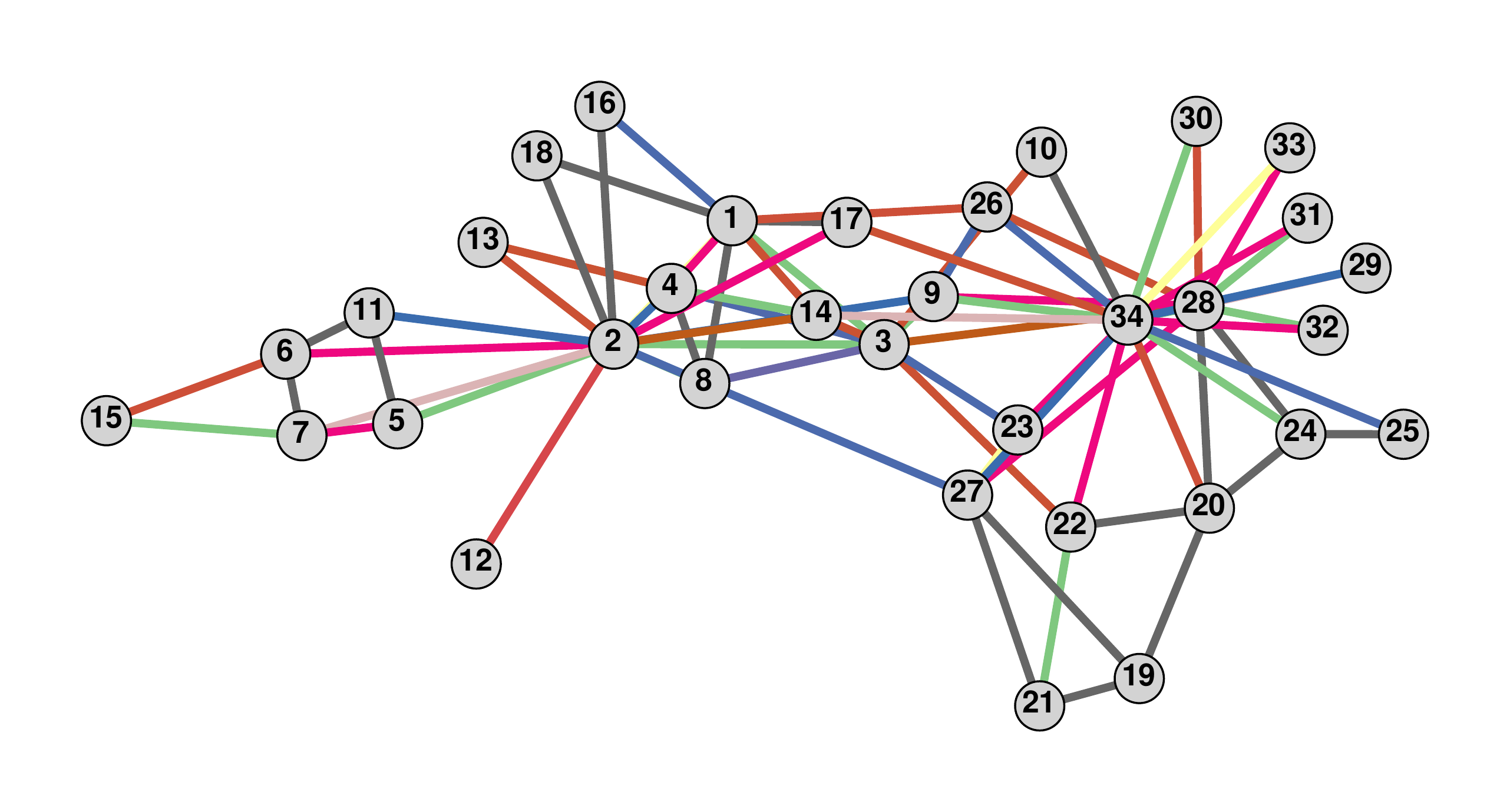}\\
\hspace{-.5cm}(a) RGB&\hspace{-.6cm}(b) LAB ($0\le L\le 70$)&\hspace{-.6cm}(c) ColorBrewer Accent\_8\\
\includegraphics[width=.35\linewidth]{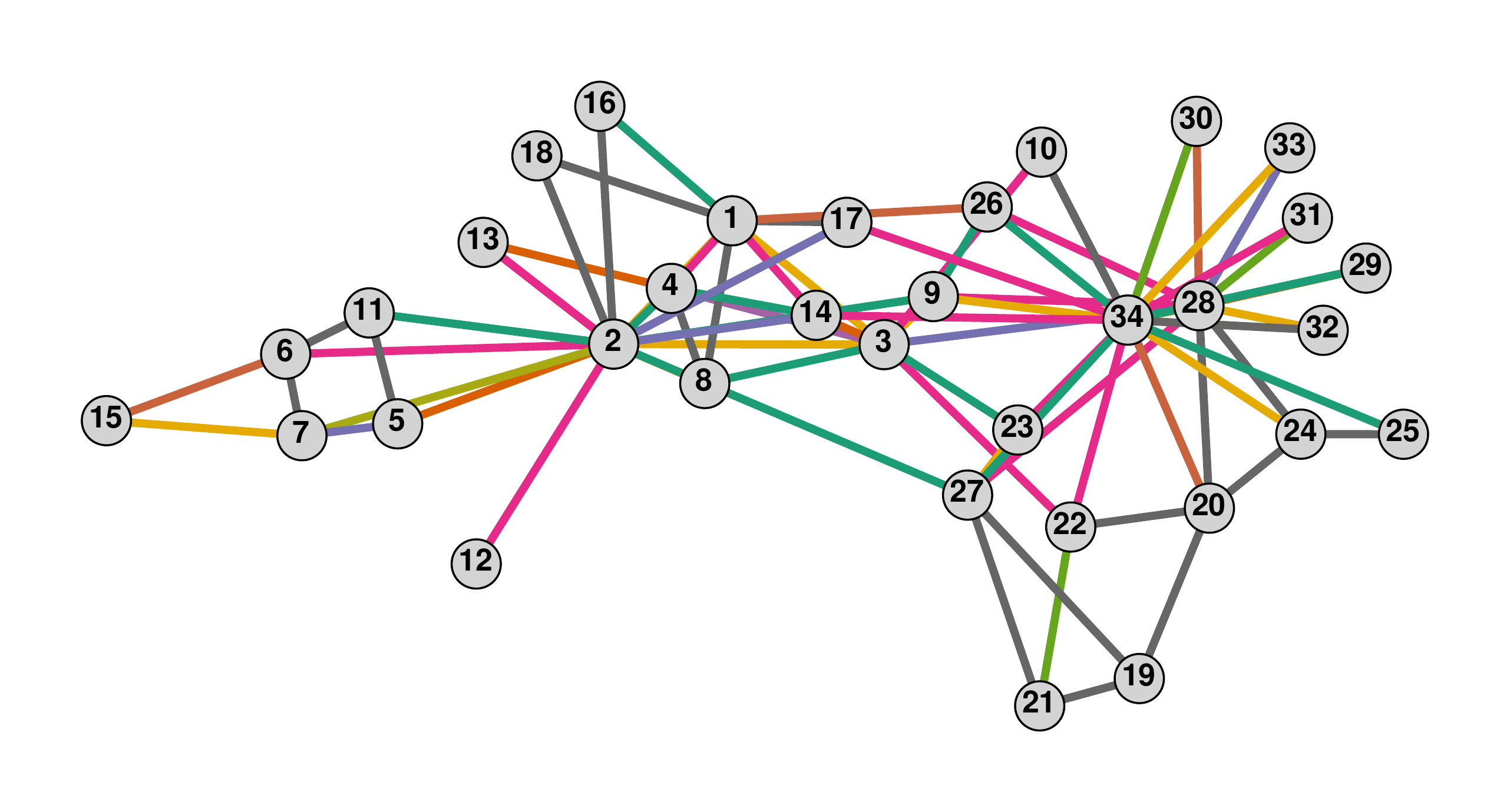}
&\hspace{-.5cm} \includegraphics[width=.35\linewidth]{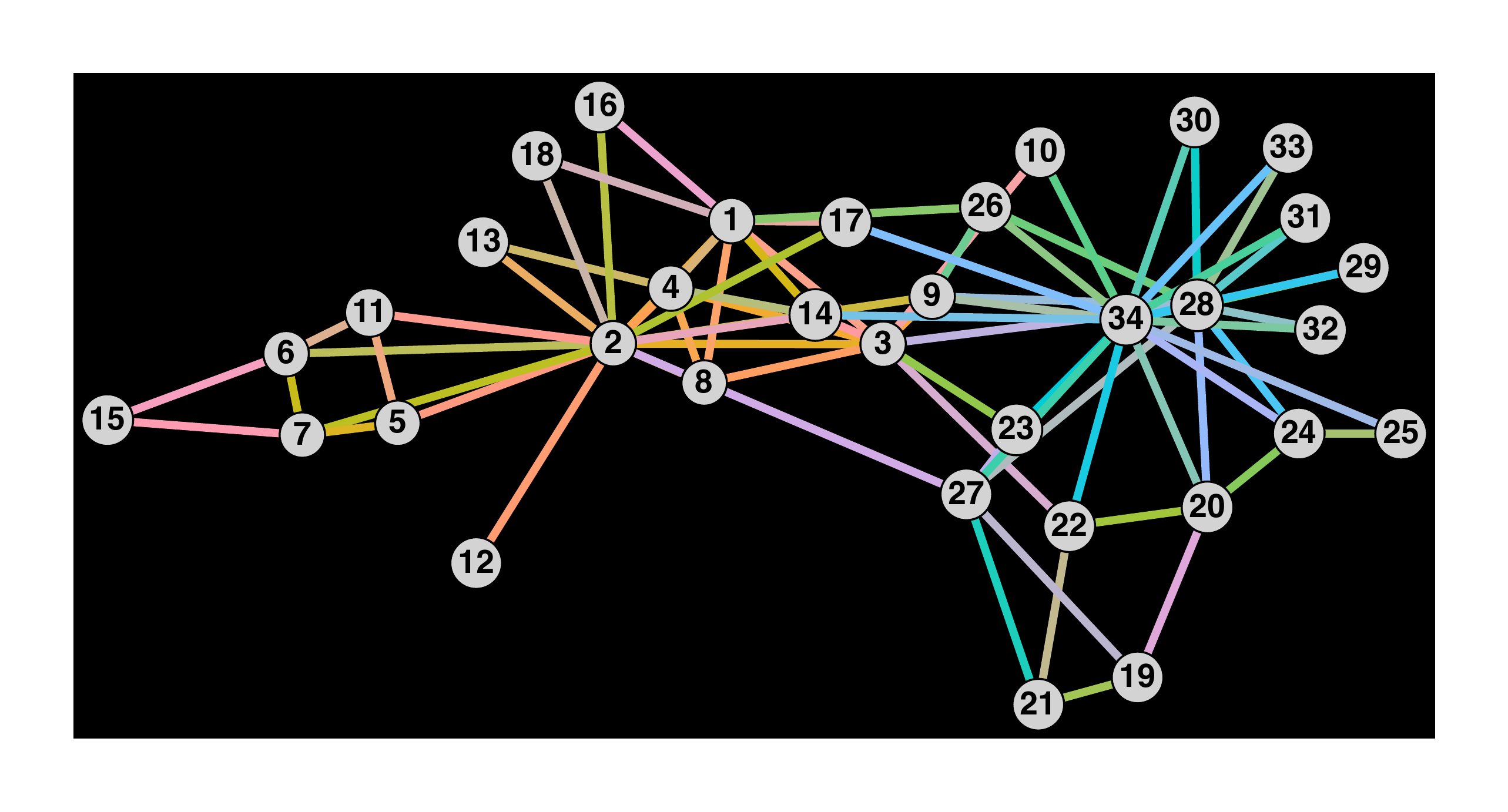}
&\hspace{-.5cm} \includegraphics[width=.35\linewidth]{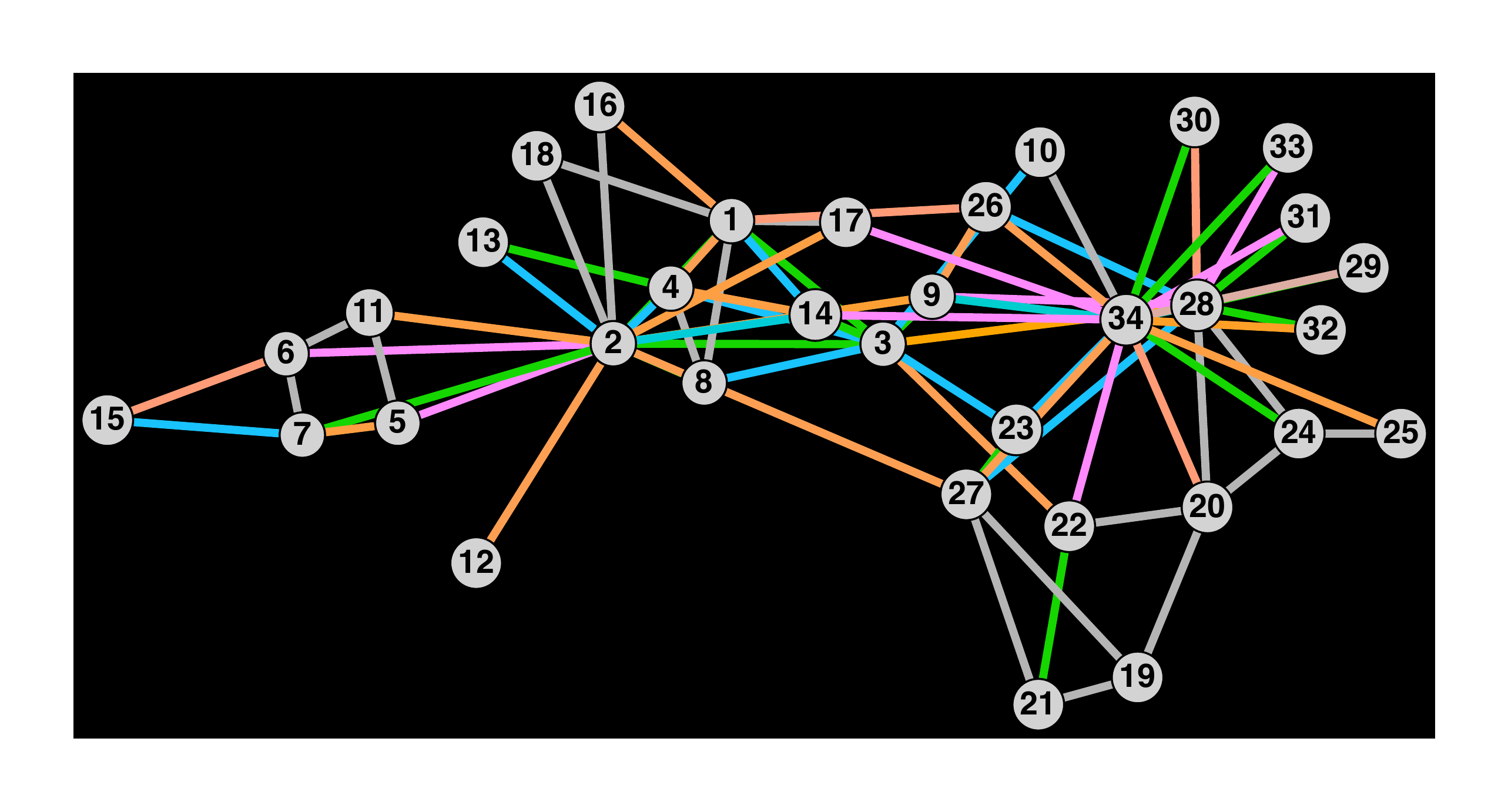}\\
\hspace{-.6cm}(d) ColorBrewer Dark2\_8
&\hspace{-.5cm}(e) applying Jianu {\textit et al.}~\cite{Jianu_2009_edge_color}
&\hspace{-.5cm}(f) LAB ($L=75$)\\
%\hspace{-.5cm}\fcolorbox{white}{darkgray}{\includegraphics[width=4.1cm]{DATA/karate_graphplot_color_ColorBrewer_pastel1_9.pdf}}
%&\hspace{-.5cm} \fcolorbox{white}{darkgray}{\includegraphics[width=4.1cm]{DATA/karate_graphplot_color_lab_l80.pdf}}\\
%(e) ColorBrewer Pastel1\_9
%&(f) LAB ($L\ge 80$)\\

\end{tabular}
\end{center}
\vspace{-.6cm}
\caption{\textsf Applying CLARIFY on the Karate graph in RGB and
  LAB color spaces (a-b), and with two ColorBrewer palettes (c-d). For comparison we
  include the result of applying the algorithm of Jianu {\textit et
    al.}~\cite{Jianu_2009_edge_color}, vs CLARIFY in LAB color
  space with fixed lightness of 75 (e-f).
%For (e) and (f), we used assume that a dark background is required. In (e)
%a light pastel color palette is used, while in (f) we allows only LAB colors with intensity $> 80$.
\label{palettes}}
\vspace{-.5cm}
\end{figure*}

Instead, because CLARIFY works just as well on a {\em discrete} set of colors, we modify
CLARIFY slightly as follows. We first sample the LAB gamut: we subdivide L, A and B
at one unit increment, and check whether the resulting points are inside the LAB gamut by converting
the point to RGB space, and back to the LAB space. If the double-conversion ends at the same point (within
a threshold of 0.02 in Euclidean distance), the point is considered inside the LAB gamut. This resulted in 826816 points (12.4\% of the LAB space).
Note that we only have to find this sample set once and store as a file.
We then construct an octree over this point set. The CLARIFY algorithm works with this
octree, without worrying about staying inside the LAB gamut. This sampling technique also makes it very easy
to control the lightness of the color -- if we need to display the drawing in a dark background and
thus light colors are desired, we can simply filter out points with a low $L$ value in the sample.
Fig.~\ref{palettes}(b) shows the result of apply CLARIFY in the LAB space with $0\le L\le 70$.
%, while
%Fig.~\ref{palettes}(f) shows the result with $L\ge 80$ in a black background.

In terms of CPU time, we found that working in LAB space with the sampling technique gives
very similar CPU time to working in the RGB space. Speed can be further improved
if we take a coarser sample.

%EDGE_COLORING_JAVA/edge_distinct_coloring ../REPORT/EDGE_COLORING/DATA/karate_graphplot.gv  | neato -Tpdf -n2 -Gbgcolor=''#000000'' -Epenwidth=2 > /tmp/karate_graphplot_jianu.pdf

\setlength\fboxsep{-5pt}
\definecolor{darkgray}{gray}{0.4}
\definecolor{white}{gray}{1}

\textbf{User-define color palettes.} Any user defined color palette can be handled in a similar way to the LAB gamut -- we convert the
color palette consists of $k$ colors to the LAB space, then interpolate these $k$ colors to get
$K$ sample points.
We do so by subdividing the path linking these $k$ points in the LAB space into $K-1$ segments of equal distance.
The path can be constructed along a natural ordering of the palette, or 
along a shortest path/tour by solving a Traveling Salesman Problem in 3D.
An octree is then constructed using the $K$ sample points and CLARIFY is applied over the octree.
Fig.~\ref{palettes} gives some examples of using two ColorBrewer~\cite{ColorBrewer2} color palettes, with $K = 10^4$.

\textbf{Other color spaces.} There are situations where
other color space may be more appropriate. For example, for disambiguating 
the edges in Fig.~\ref{dual}, we avoided using colors for edges in order to 
accurately display colors of the nodes. For this drawing we used CLARIFY with the gray scale, so that edges are
in black or gray. We could even map the gray scale to
line styles, with black for solid line and gray for dashed line of different weights.

%On larger graphs, 
%the algorithm often do not converging.
%We dod not comparing color due to the
%different objective function, and that they limit to 2D.

\begin{comment} Timing on a mac
problem Jianu CLARIFY
ngk 1.8 0.6
Notra 25 1.1
GD 11 1
erdos 15 1.4
harvard 30 4
\end{comment}

\subsection{Complexity of the CLARIFY algorithm}

The CLARIFY algorithm consists of two main steps: finding the dual collision graph, and computing a color
assignment. 

The dual graph is calculated by checking whether edge pairs are in collision. Conditions C2 and C3 
can be checked by looping through each node of the original graph, and testing if
a pair of edges starting from the node nearly overlap, or run in almost opposite directions.
This check can be done after sorting the angles, hence on a node with $d$ neighbors, assuming that the edges are not 
entirely on top of each other, the cost should be around $d \log(d)$, so the 
cost of checking over all nodes is $|E|\log|E|$ (the pathological case of all edges on top of each other
would give a complete dual graph thus a complexity of $|E|^2$).
% In our implementation we make checking
%of C3 optional, because if edges are allowed to be drawn on top of nodes, then edges starting from a node set off at the
%border of the node, while edge passing through a node are drawn on top of the node, thus C3 would not cause any confusion.

Condition C1 can be checked using the Bentley-Ottmann algorithm~\cite{ORourke_1998_compgeom} with a complexity of $O((|E| + k) \log |E|)$, where $k$ is the number of edge crossings. If $k$ is $|E|^2$ or more, a naive algorithm which checks all $|E|^2/2$ edges should be used. We are not aware of a good algorithm for checking C4, one possibility is to replace each edge with 
a rectangle in the shape of a thicker edge, then apply the Bentley-Ottmann algorithm, which should give us the same
complexity as checking C1.

The second step of CLARIFY, that of assigning colors, applies the EmbedOneNode algorithm repeatedly 
over all nodes. EmbedOneNode is a branch-and-bound algorithm over an octree data structure. Its complexity
is dependent on the number of local maxima, and how close they are to the global maximum (in terms of the objective function value). If the local
maxima have much smaller function values compared with the global maximum, as in the case of Fig.~\ref{contour}, then
 branches of the octree/quadtree corresponding to the local maxima
 will terminate at an early stage, and the
complexity of the algorithm is around $\log(\epsilon)$, otherwise the complexity is around $L*\log(\epsilon)$ where
$L$ is the average number of local maxima. Overall the worst case complexity is $O(|E|\log(\epsilon) L)$
per iteration over all nodes. $L$ is a value hard to quantify, we believe it is related to the 
average degree of the dual graph.

Taking both the collision graph formation and the optimization into account,  the CLARIFY algorithm has an average case complexity of $O((|E| + k) \log |E| + t |E|\log(\epsilon) L)$, with $k$ the number of edge crossing, $L$ the average number 
of local maxima, and $t$ the number of iterations of Algorithm~1.
The worst case complexity is $O(|E|^2 + t |E|\log(\epsilon) L)$, in the pathological situation where all edges are on top of each other.

In practice we found that the optimization step dominates the computation time even when we
use the naive algorithm for computing the collision graph (see Table~\ref{timing}). Therefore
for the rest of the paper we use the naive algorithm for the first step of forming the 
collision graph, which makes computation of C4 much simpler.

\begin{comment}
\begin{table}[h]
\caption{Effect of $\epsilon$ on the color difference and CPU time when applying CLARIFY (ten random starts) on the graph in Fig.~\ref{random_graph}. The CPU time is that for CLARIFY, minus the time for constructing the dual graph (0.04 seconds). The latter is independent of $\epsilon$\label{epsilon}}
\begin{center}
\begin{tabular}{cccc}
\hline
$\epsilon$ & cdiff & CPU\\
\hline
$10^{-1}$& 0.866 &0.02\\
$10^{-2}$& 0.974 &0.05\\
$10^{-3}$& 0.988  &0.09\\
$10^{-4}$&  0.990 & 0.23\\
$10^{-5}$& 0.990  & 0.43\\
\hline
\end{tabular}
\end{center}
\end{table}
\end{comment}

\subsection{Choice of parameters}

For checking collision conditions, we need to define what is a ``small
angle'' and what is ``close to 180 degree.'' Based on empirical
observations, by default we set these to be $15$ degree and $165$
degree. We define two lines being ``very close'' if their distance
(the smallest distance between two points on the lines) is less than
1\% of the larger of the length of the lines. We consider two lines as
``almost parallel'' if they form an angle that is less than one
degree.  The parameter $\epsilon$ controls the accuracy with which we
find the global optimal embedding for one node. 
%Table~\ref{epsilon}
%shows the effect of this parameter on the overall color difference
%achieved, as well as on CPU time. Clearly
Through numerical experiment we found that
 the CPU time increases
almost linearly with $\log(\epsilon)$, as predicted by the complexity
analysis. The color difference %(cdiff) 
achieved is also in-line with
expectation: from $\epsilon$ to $\epsilon/10$, %cdiff 
it changes roughly
proportionally to $d^{1/2} \epsilon$ or less, where $d=3$ is the
dimension of the color space. This fits our analysis in
Section~\ref{color_opt}. Perceptually, we found that $\epsilon =
10^{-2}$ gives very similar coloring to $\epsilon=10^{-3}$, hence we
set $\epsilon=10^{-2}$ by default.

\begin{figure*}[htbp]
\vspace{-.5cm}
\def\tabcolsep{1pt}
\begin{center}
\begin{tabular}{cc}
\includegraphics[width=.5\linewidth]{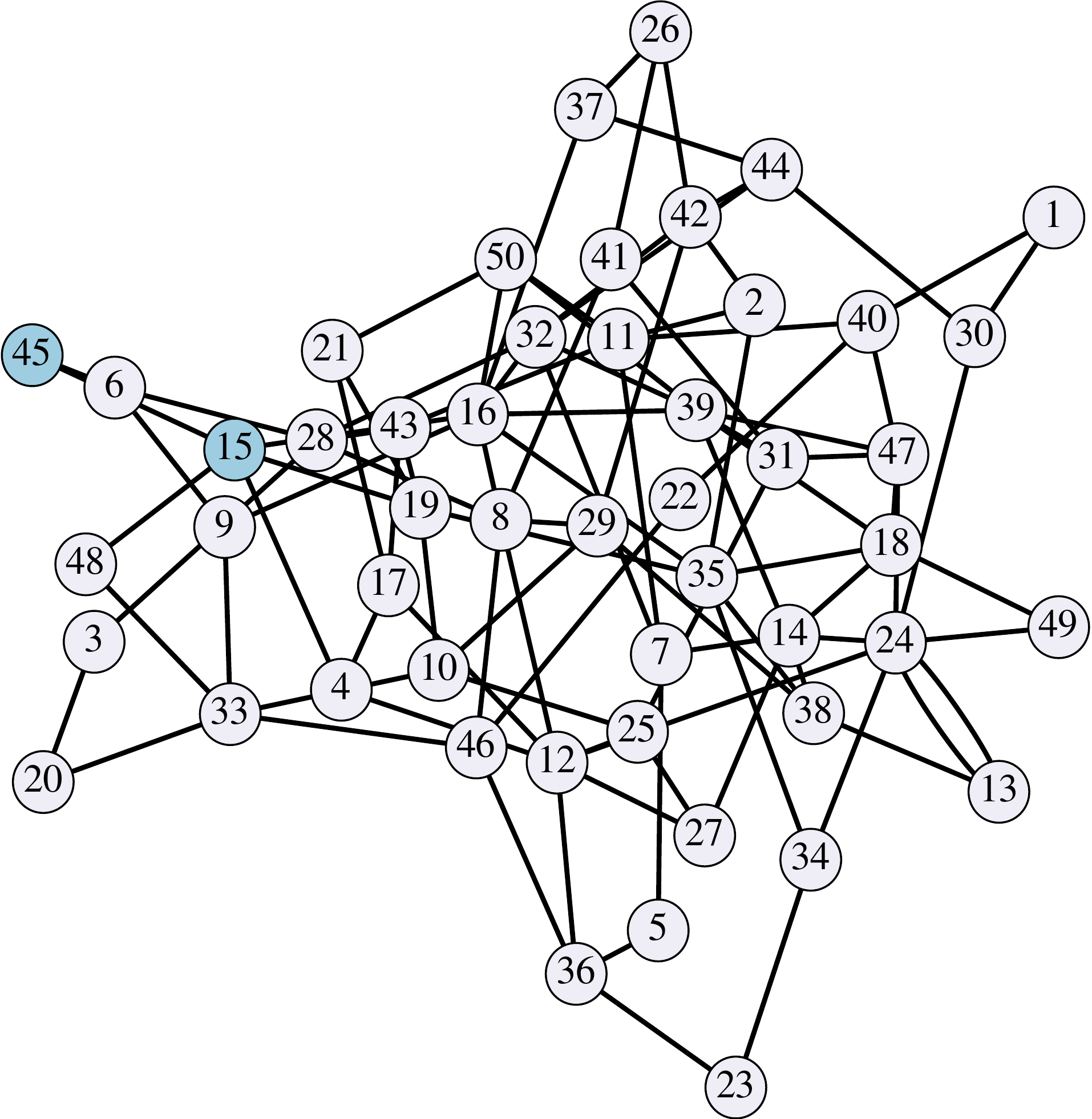}&\hspace{-.05cm}\includegraphics[width=.5\linewidth]{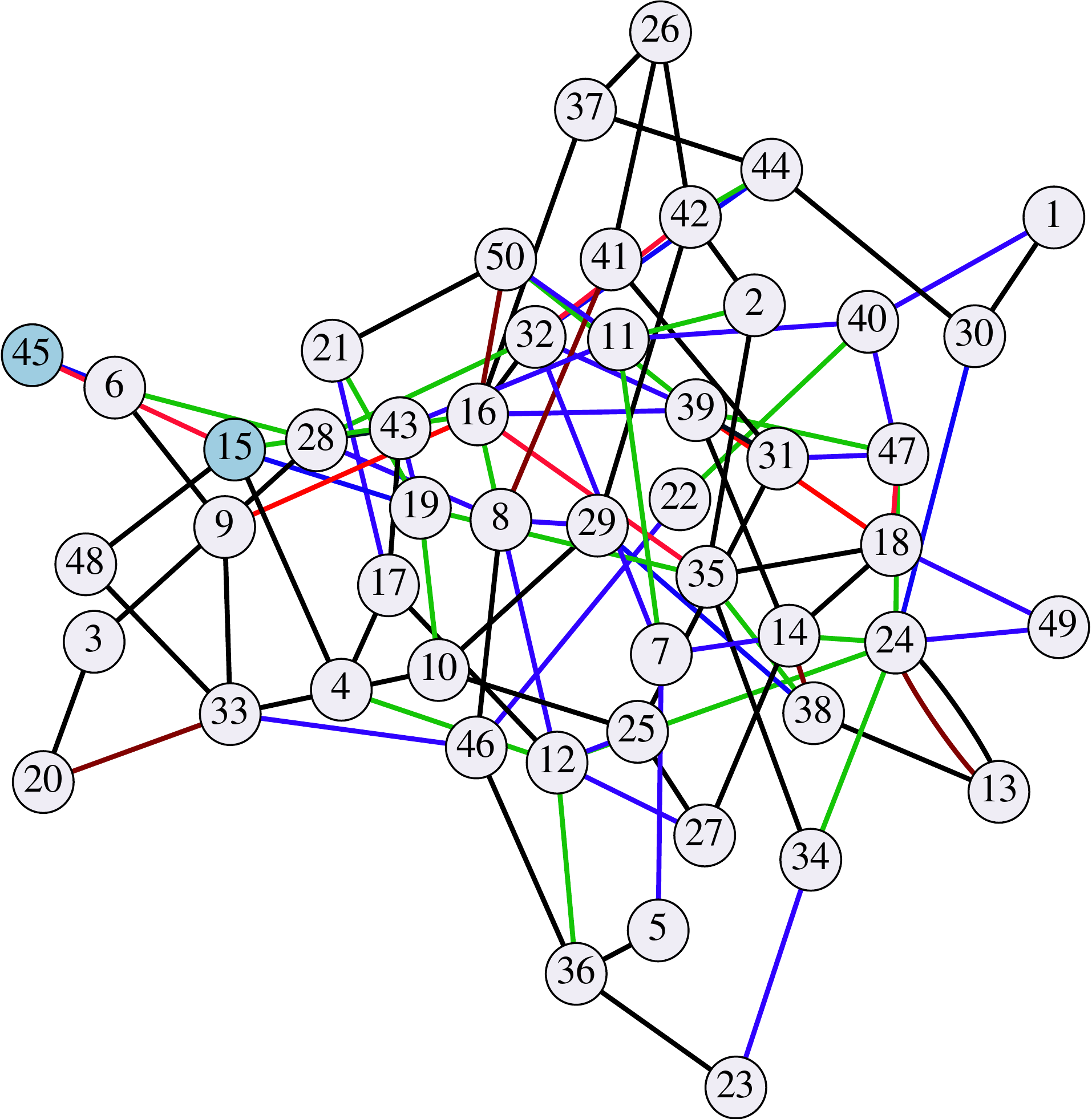}\\
[-.6cm]\\
(a)&(b)\\
\end{tabular}
\end{center}
\vspace{-.5cm}
\caption{\textsf Edge coloring on ngk\_4 graph: (a) the original graph. Are nodes 45 and 15 (blue) connected? (b) the colored drawing. We can tell that 45 and 15 are indeed connected by a red edge.\label{ngk}}
\end{figure*}

\begin{figure*}[htbp]
\vspace{-.2cm}
%\hspace{-.5cm}\hbox{\includegraphics[width=.42\linewidth]{DATA/custa5.pdf}\includegraphics[width=.1\linewidth]{DATA/custa5_zoom.jpg}\hspace{-.3cm}\includegraphics[width=.42\linewidth]{DATA/custa5_color.pdf}\includegraphics[width=.1\linewidth]{DATA/custa5_color_zoom.jpg}}
\begin{tabular}{cccc}
\hspace{-.5cm}\includegraphics[width=.42\linewidth]{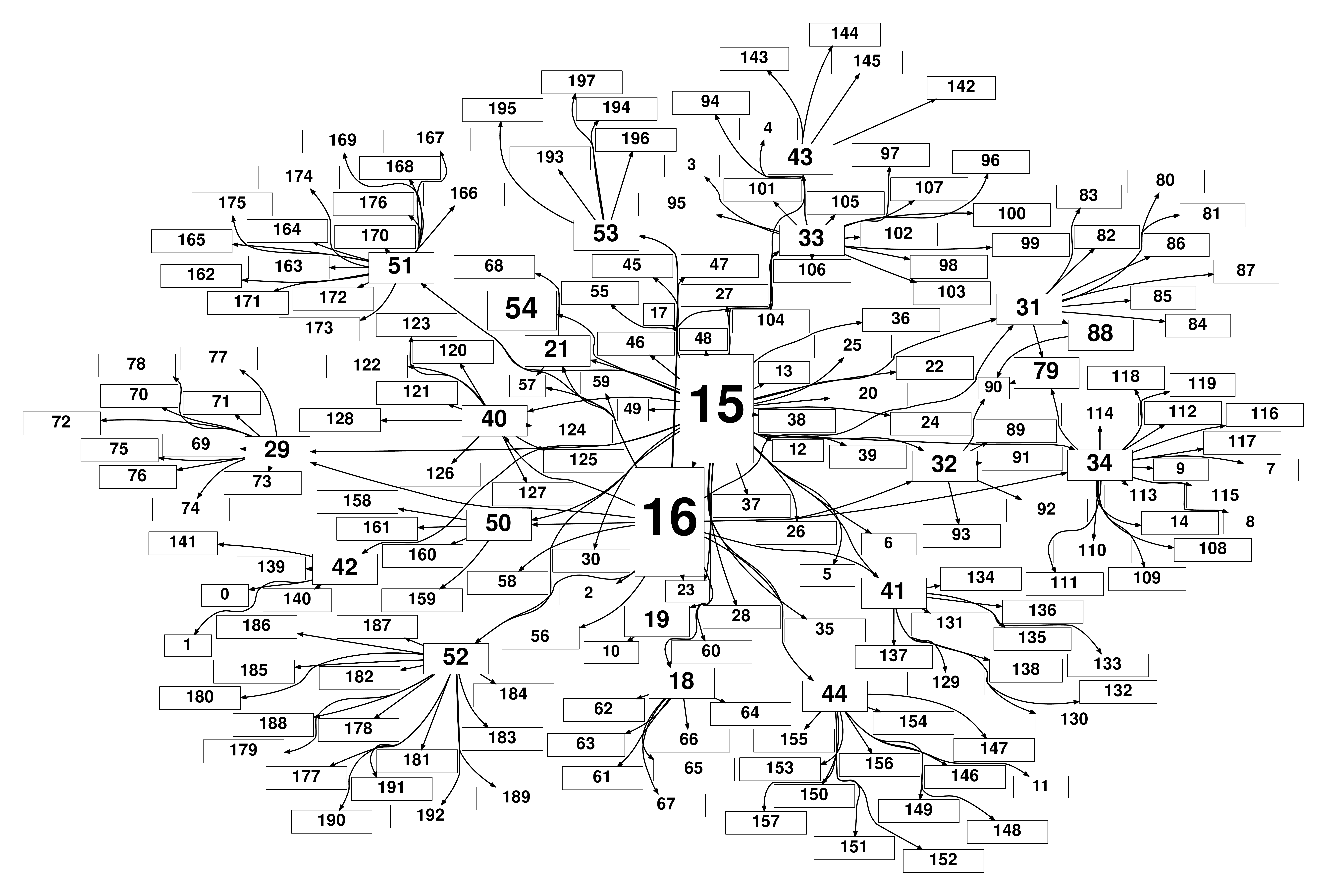}&
\hspace{-.5cm}\includegraphics[width=.1\linewidth]{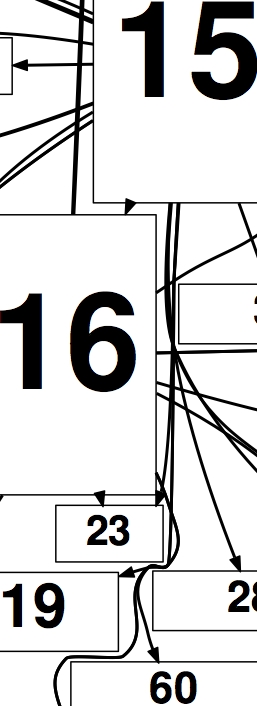}&
\hspace{-.6cm}\includegraphics[width=.42\linewidth]{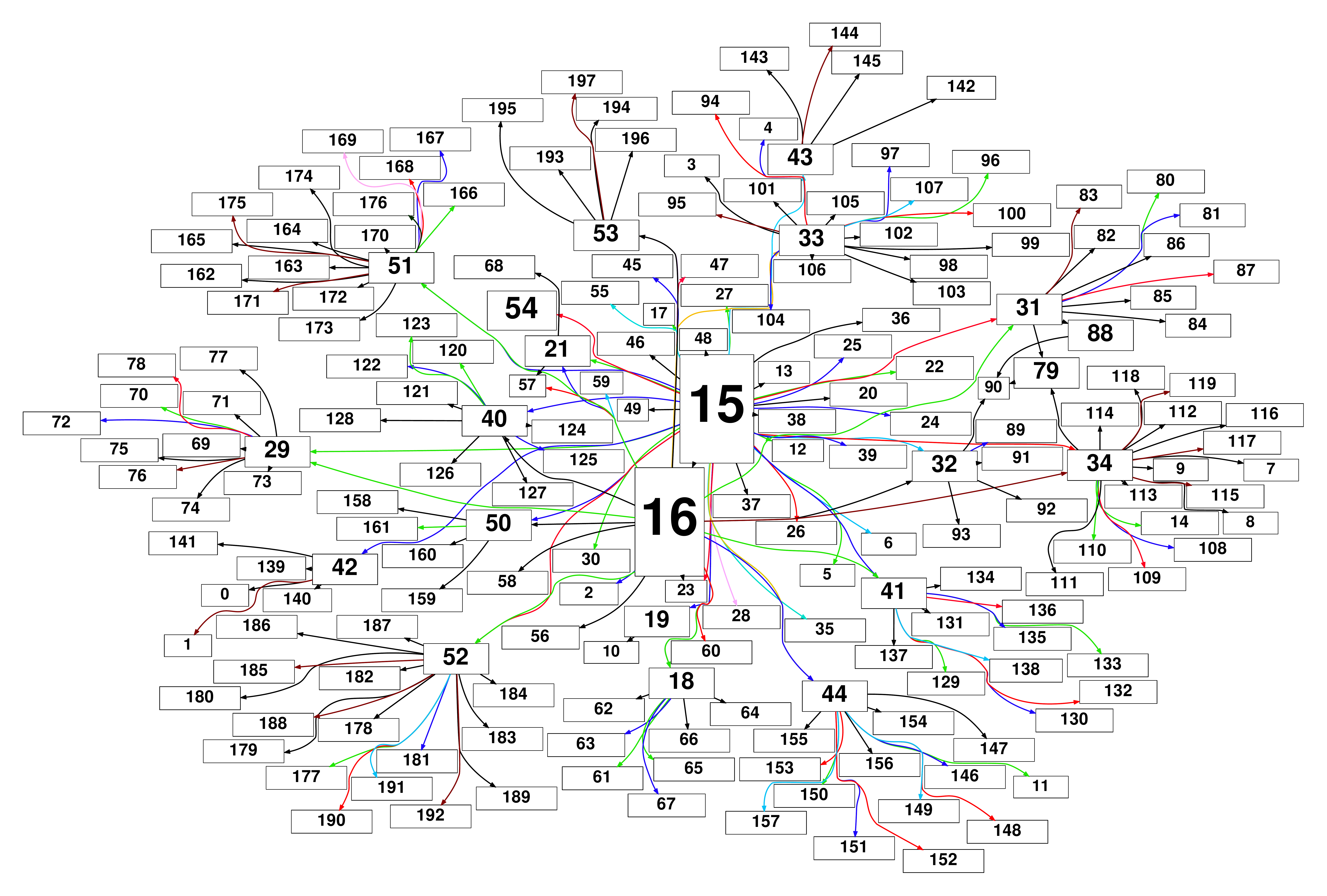}&
\hspace{-.5cm}\includegraphics[width=.1\linewidth]{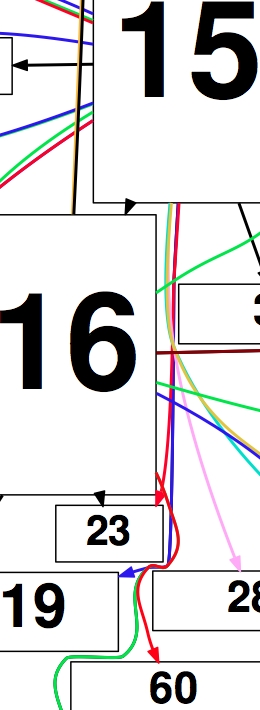}\\
[-.5cm]\\
(a)&(b)&(c)&(d)\\
\end{tabular}
\vspace{-.4cm}
\caption{\textsf (a) A graph with spline edges. Some of the splines are hard to differentiate. (b) In the zoomed-in view, is node 16 connected to node 60, or to node 19 (both below node 16)? (c) Splines are colored using the CLARIFY algorithm. Now colliding
edges are easier to differentiate. (d) In the zoomed-in colored view, node 16 is seen to be
connected to node 60 by a red spline, but not to 19. The latter is connected by a blue spline to node 15 above.\label{custa}}
\vspace{-.3cm}

\end{figure*}

\begin{figure*}[htbp]
\hspace{-1.5cm}
\begin{center}
\begin{tabular}{cccc}
\hspace{-.2cm}\includegraphics[width=.5\linewidth]{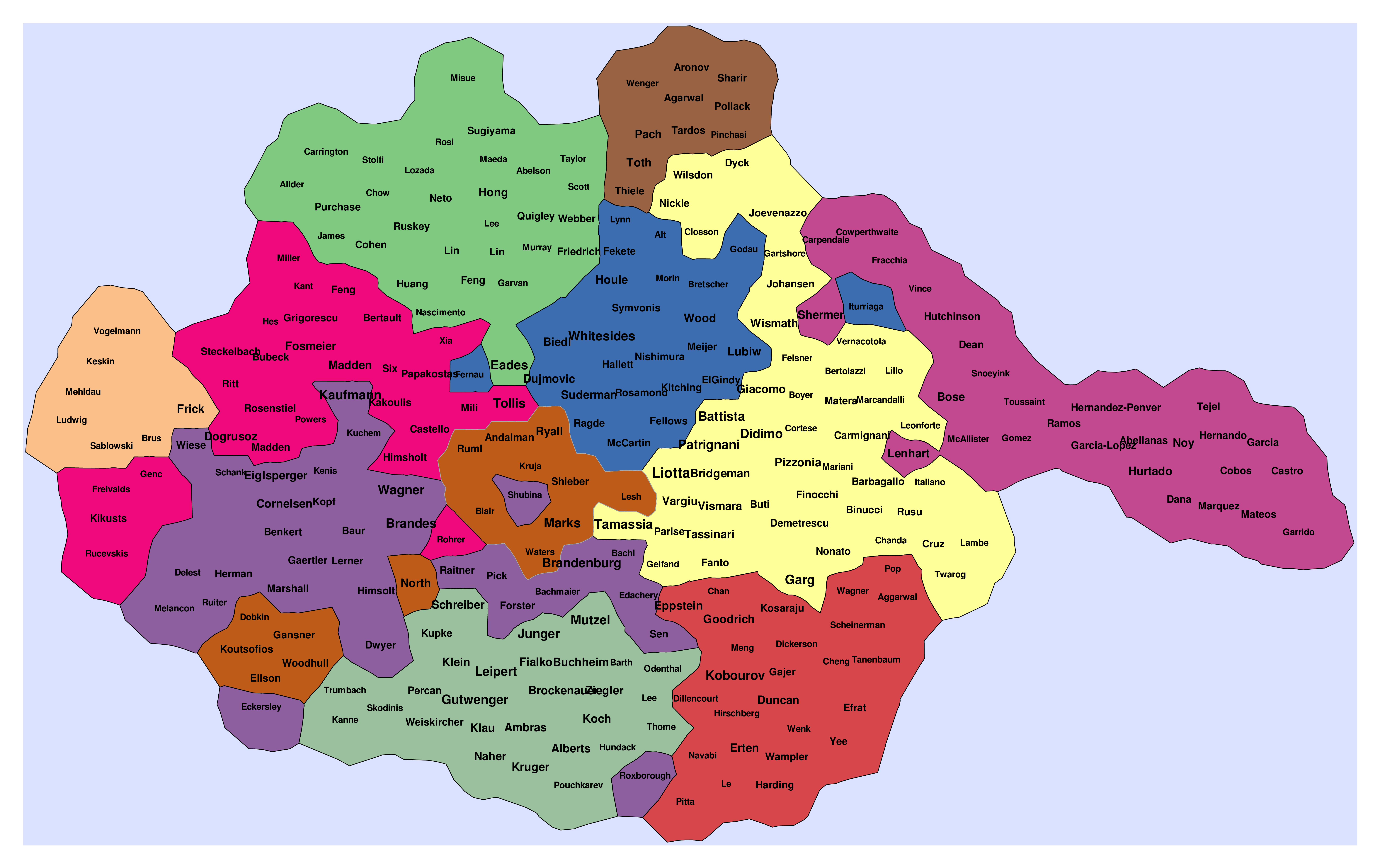}
&\hspace{-.6cm} \includegraphics[width=.5\linewidth]{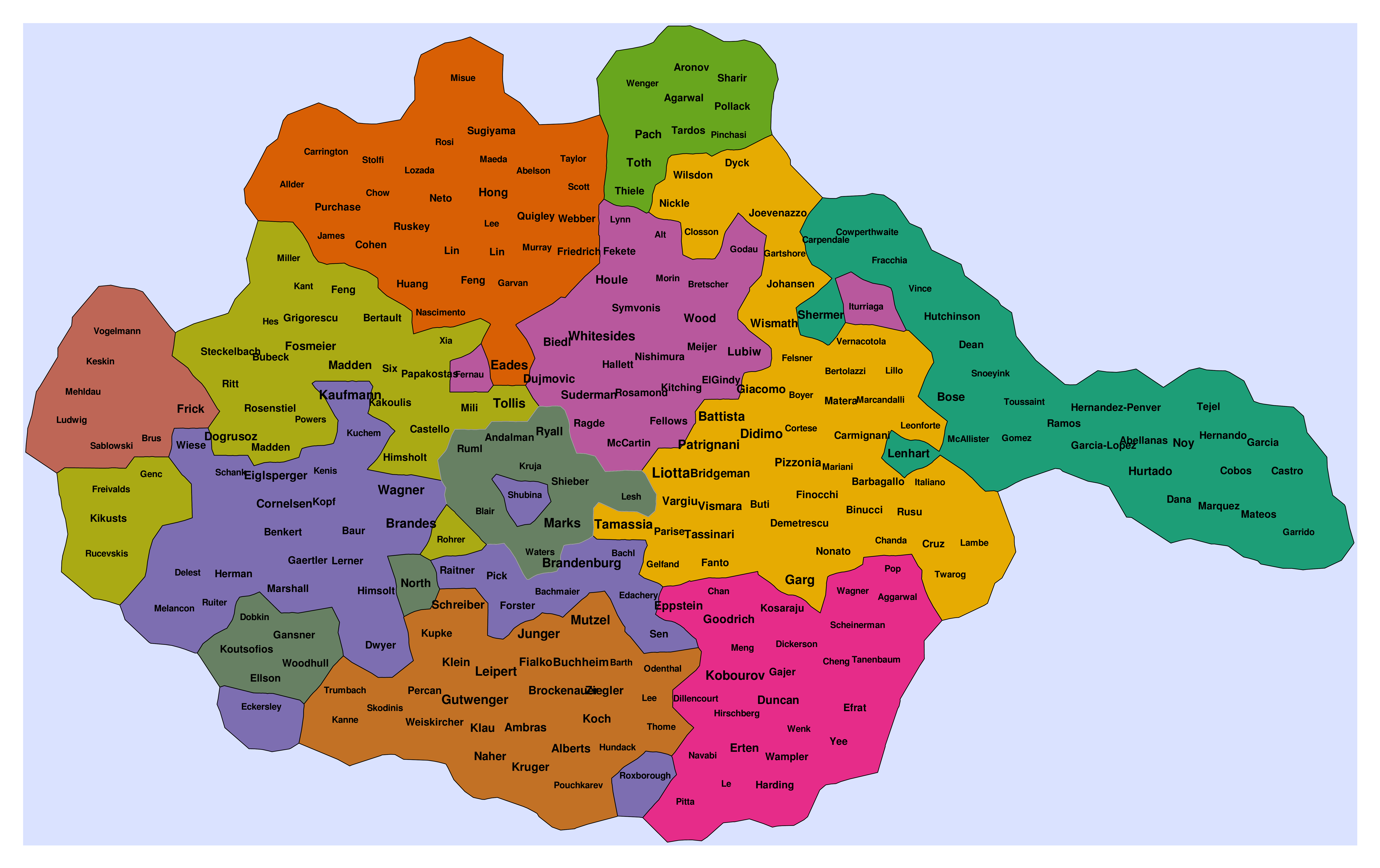}\\
%\vspace{-.5cm}\\
%\hspace{-.2cm}\includegraphics[width=.5\linewidth]{DATA/COLOR_PALETTES/gd_1994_2004_sfdp_1comp_ColorBrewer_brbg_9.pdf}
%&\hspace{-.6cm}\includegraphics[width=.5\linewidth]{DATA/COLOR_PALETTES/gd_1994_2004_sfdp_1comp_ColorBrewer_pastel1_9.pdf}
%(a) Accent\_8
%&(b) Dark2\_8\\
%&(c) Pastel1\_9
%&(d) \cite{Hu_2010_gmap}\\
\end{tabular}
\end{center}
\vspace{-.5cm}
\caption{\textsf Applying CLARIFY on a collaboration map with two ColorBrewer palettes: left: Accent\_8, right: Dark2\_8.  \label{maps}}
\vspace{-.5cm}
\end{figure*}

\subsection{Examples}

We now apply CLARIFY to graphs from real
applications. Table~\ref{timing} gives results on six of the graphs
we tested, including running time and objective
function~(\ref{optimization}) (color diff) achieved in LAB color space.
These come either from the University of Florida Sparse
Matrix Collection~\cite{Davis_Hu_ufl_2009}, or from the test graphs distributed with Graphviz~\cite{graphviz:2000},
and originate from
different application areas. 
We intentionally avoided choosing mesh-like
graphs -- such graphs %have a low complexity and 
are
 easy to layout aesthetically. Their layouts also tend to exhibit
a low perceptual complexity, making it relatively easy to 
follow edges and paths. Compared with a non-mesh-like graph, a
mesh-like graph is easier for our algorithm 
because there are typically fewer colliding edges.
We ran the experiment on a Macbook Pro laptop with a 2.3 GHz Intel Core i7 processor.

\begin{table}
\vspace{-.1cm}
\caption{Statistics on the original and dual test graphs, CPU time (in
  second) and objective function (cdiff) for CLARIFY (one random start). The time in bracket is for constructing the dual collision graph.\label{timing}}
\vspace{-.5cm}
\begin{center}
\begin{tabular}{ccccccc}
\hline
graph &  $|V|$ & $|E|$ & $|E_c|$ & CPU & {\text cdiff}\\
\hline
ngk\_4          & 50 & 100 & 54 & 0.6 (0.) & 122.69\\
NotreDame\_yeast & 1458 & 1948 & 1685 & 1.3 (0.2) & 67.9\\
GD00\_c &  638 & 1020 & 1847 & 1.7 (0.1) & 64.32\\
Erdos971 & 429 & 1312 & 4427 & 2.1 (0.1) & 59.3 \\
Harvard500 & 500 & 2043 & 11972 & 2.3 (0.3) & 35.0\\
extr1     & 5670 & 11405  & 34696 & 14.5 (7.9) & 47.1\\
\hline
\end{tabular}
\end{center}
\vspace{-.6cm}
\end{table}

\begin{comment}% linux verus machine
ngk\_4          & 50 & 100 & 54 & 0.1 (0.1)\\%colordiff  0.859760
NotreDame\_yeast & 1458 & 1948 & 1685 & 1.1 (0.4)\\ %0.538
GD00\_c &  638 & 1020 & 1847 & 0.8 (0.2)\\ %0.299 color diff
Erdos971 & 429 & 1312 & 4427 & 2.6 (0.2)\\ %0.427 color diff
Harvard500 & 500 & 2043 & 11972 & 5.9 (0.6)\\ %0.481 color diff
\end{comment}

It can be seen from Table~\ref{timing} that for graphs of up to a few
thousand nodes and edges, CLARIFY runs quickly. The majority of the
CPU time is spent on color assignment, while the construction of the
dual graph takes relatively little time even with the naive dual
graph construction algorithm. 
%Overall, the CPU time seems to be 
%linearly proportional to the number of edges in the dual graph.
The Harvard500 graph gives a large $|E_c|$ (number of edges in the
dual graph) in comparison to the number of edges, because 
it has a few  almost complete subgraphs, which
results in a lot of crossings at small angles.

Fig.~\ref{ngk} shows the ngk\_4 graph before and after the coloring.
It is difficult to tell, from Fig.~\ref{ngk}(a), whether nodes 45 and 15 (blue) are connected.
From Fig.~\ref{ngk}(b) we can tell that they are indeed connected by a red edge.

So far we have been applying CLARIFY to straight-line drawings of graphs. The algorithm can also be used for 
drawings where edges are splines. This could be the result of an edge bundling, or an edge routing. 
Fig.~\ref{custa} shows the result of applying our algorithm to a graph from a user of our software, this is one of the examples
that motivates our work. As we can see, from the original drawing, it is difficult to
differentiate some of the splines. For example, is node 16 connected to node 60, or to node 19 (both below node 16)? 
With colored splines, we can see that node 16 is connected to node 60
by a red spline. %, while 
%node 15 is connected to node 19 by a blue spline.

%slines, we currently detect cross using the polyline connecting the control points. splines. alpha = 30.

%\begin{figure}[htbp]
%\begin{center}
%\includegraphics[width=.48\linewidth]{DATA/polylines.pdf}\hspace{.1cm}\includegraphics[width=.48\linewidth]{DATA/splines.pdf}
%\end{center}
%\caption{\textsf Left: polylines connecting control points of two splines, intersect at a not-so-small angle.
%Right: corresponding spline that intersect at a much smaller angle.\label{polylines}}
%\end{figure}

Finally, we applied CLARIFY to color virtual maps where countries
could be fragmented. Because of the fragmentation, we have to use as
many color as there are countries. 
Fig.~\ref{maps} shows colored versions of 
an author collaboration map (see \cite{Hu_2010_gmap})
using two color palettes. 
Here each node is an author who published
in the International Symposium of Graph Drawing between 1994 to
2004. Authors are connected by edges if they co-authored a paper. This
gives a collaboration graph. Nodes are then clustered to form
countries. 
Up to now, for coloring edges of node-link graphs, we assume that
it is equally important to differentiate all colliding edge pairs, thus set the
$w_{ij}$ in (\ref{optimization}) to 1. 
For coloring virtual maps, it
is more important to color adjacent countries with more distinct
colors, at the same time, we also want to differentiate all countries.
Thus we set $w_{ij}$ to be the inverse of the length of the shortest
path that connect countries $i$ and $j$ in the dual graph of the map.
From Fig.~\ref{maps}
we can see that CLARIFY works well in using the specified palettes,
keeping neighboring countries colored with very distinct colors. Unlike
the coloring algorithm in  \cite{Hu_2010_gmap}, we 
also maintain good color distinction among non-neighboring countries.
Additional examples of graph and map coloring can be found in
the supplemental pdf file.

\subsection{Comparison with Jianu {\textit et
    al.}~\cite{Jianu_2009_edge_color}} 
We evaluated our algorithm against that of \cite{Jianu_2009_edge_color}
(hereafter called JRFL), using
the code kindly supplied by the authors.
Fig.~\ref{palettes}(e) gives the
result of applying JRFL on the Zachary graph.
Following \cite{Jianu_2009_edge_color}, we use a black background, because
the code sets lightness to 75. It is seen that near nodes 34 and 28, it
is difficult to differentiate edges. E.g., it is not clear whether
node 34 is connected to 27 or not, due to the colors of edges 34--27
and 34--23 being very similar. For a like-for-like comparison
Fig.~\ref{palettes}(f) is the results of
CLARIFY with fixed lightness of 75. Despite of the restricted lightness,
it does not suffer from the ambiguity seen in Fig.~\ref{palettes}(e).
We also compared with JRFL
on other graphs, and found CLARIFY
better both in terms of ability to disambiguate drawings, and
in speed. On most graphs, CLARIFY is about 10
times faster. Quantitatively, we found that JRFL always gives
much worse (smaller) color differences among colliding edges than CLARIFY, even if we
restrict lightness to 75 in CLARIFY.

\section{User Study\label{sec_user_study}}

\begin{figure}
\begin{tabular}{cc}
\hspace{-.5cm}\includegraphics[width=0.25\textwidth]{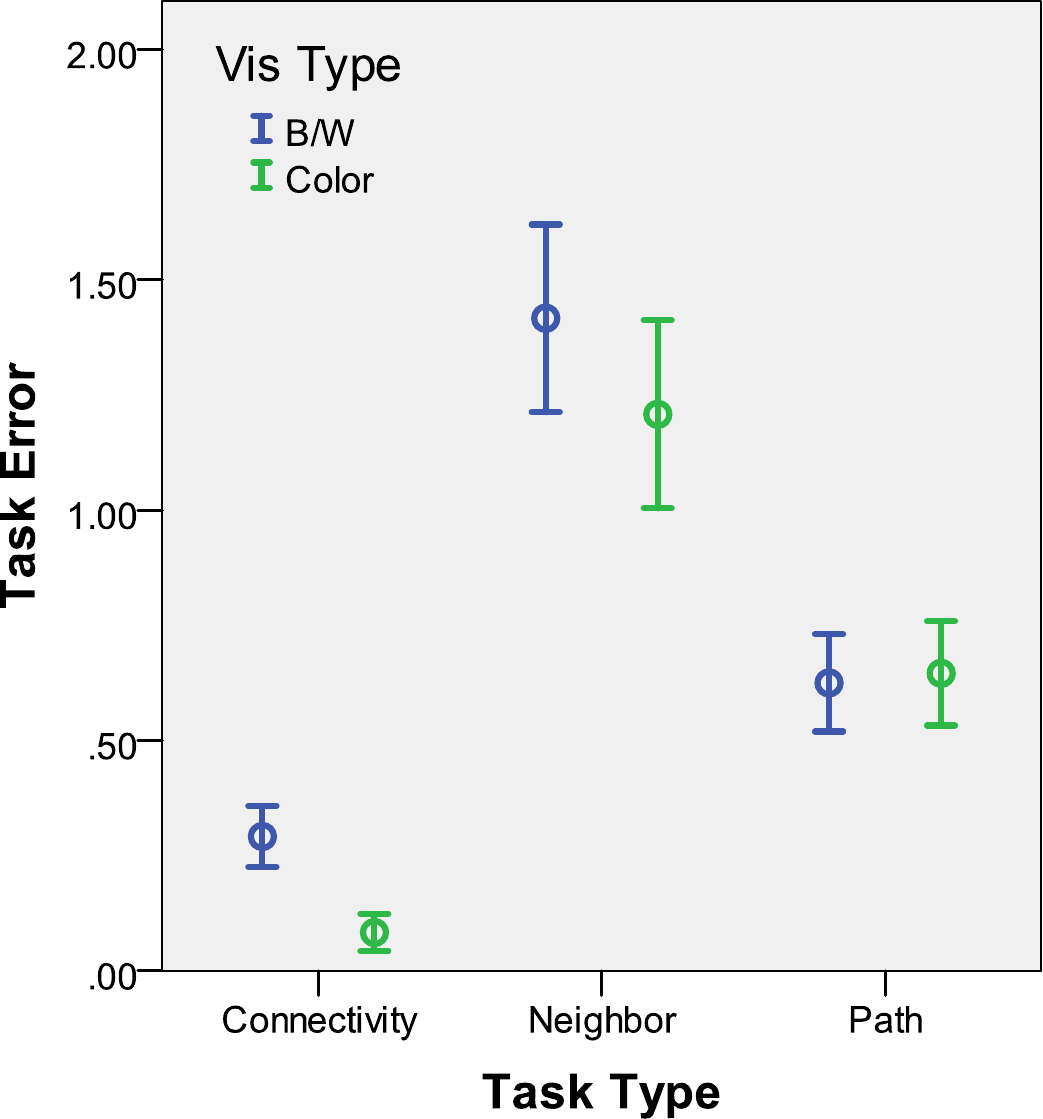}&
\hspace{-.25cm}\includegraphics[width=0.25\textwidth]{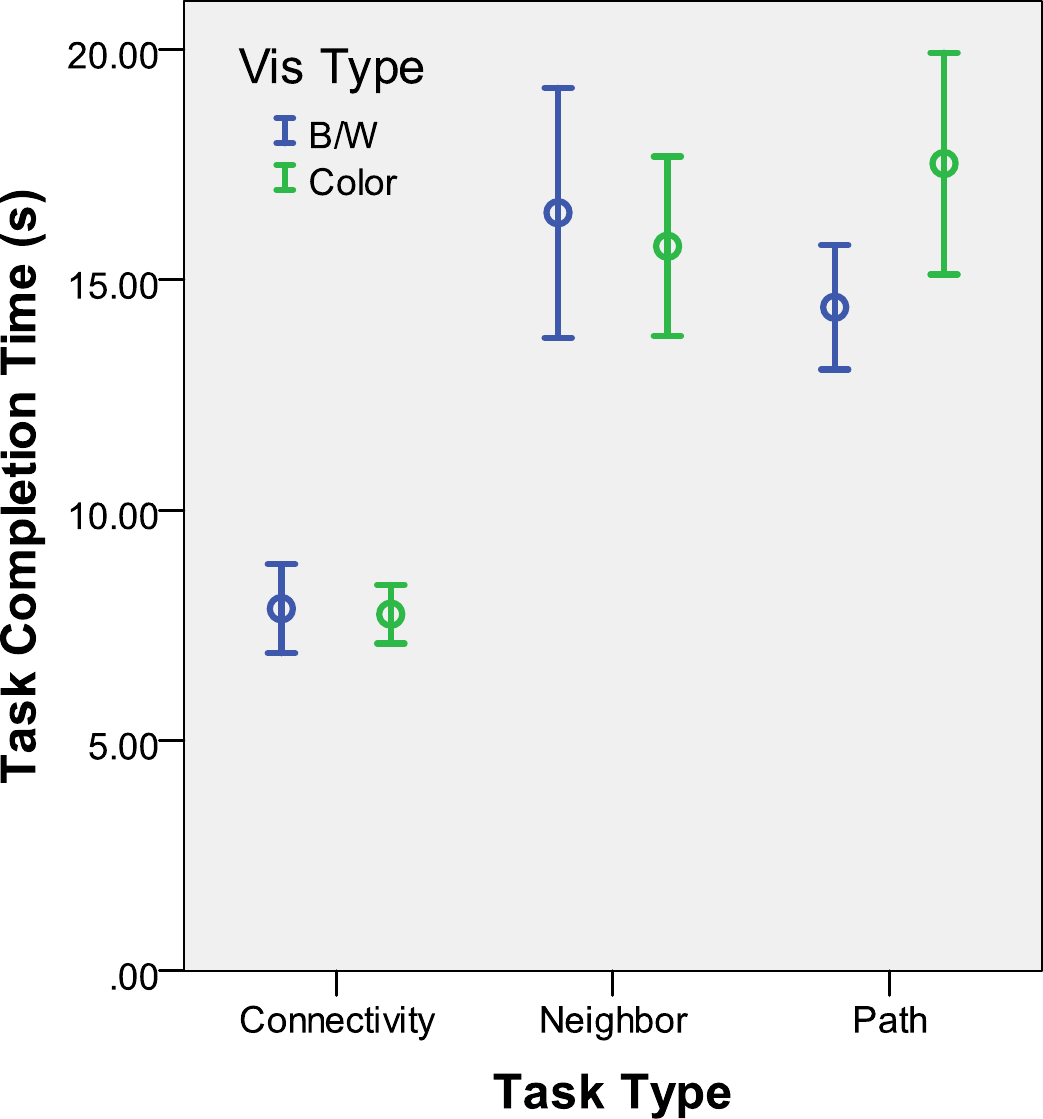}\\
(a)&(b)\\
\hspace{-.5cm}\includegraphics[width=0.25\textwidth]{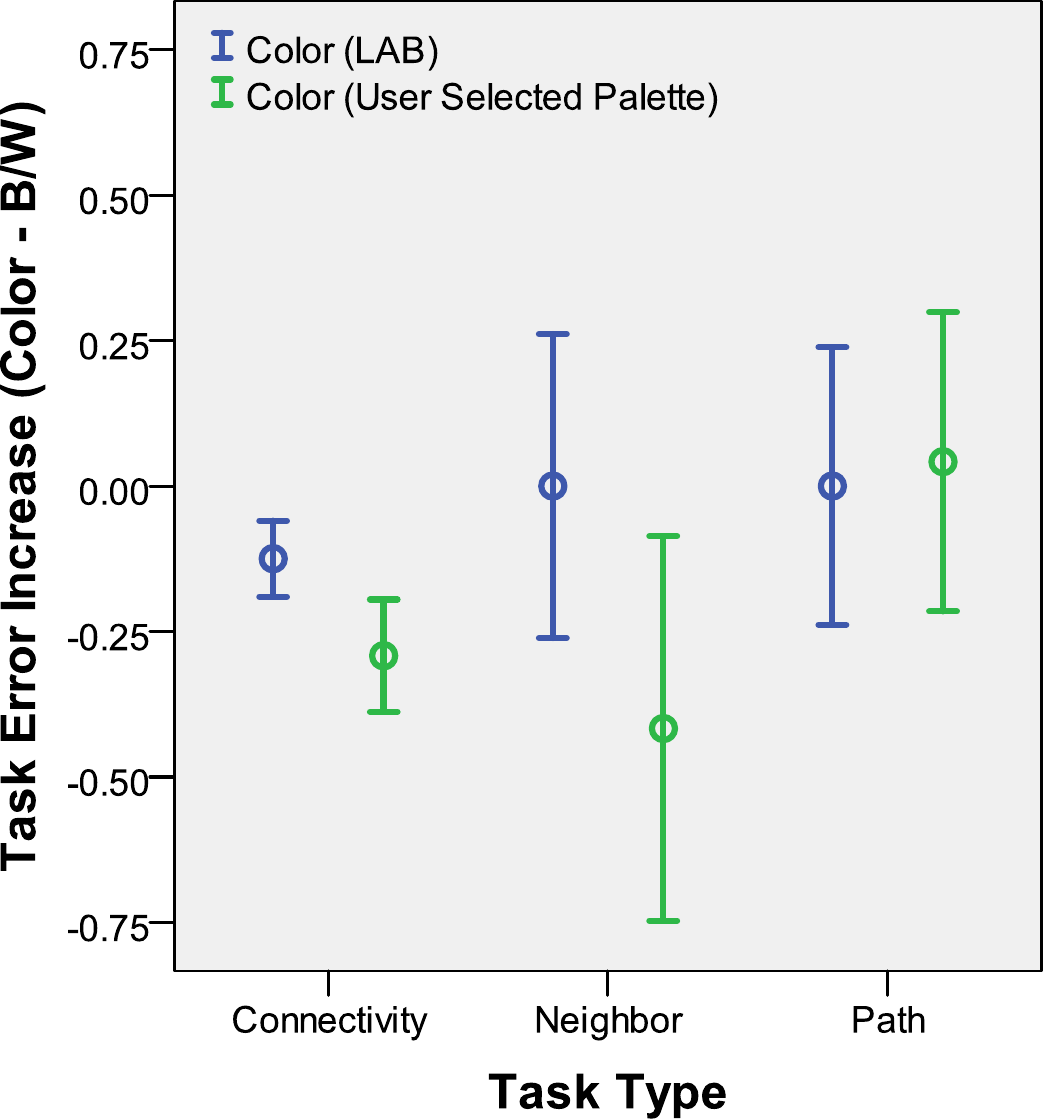}&
\hspace{-.25cm}\includegraphics[width=0.24\textwidth]{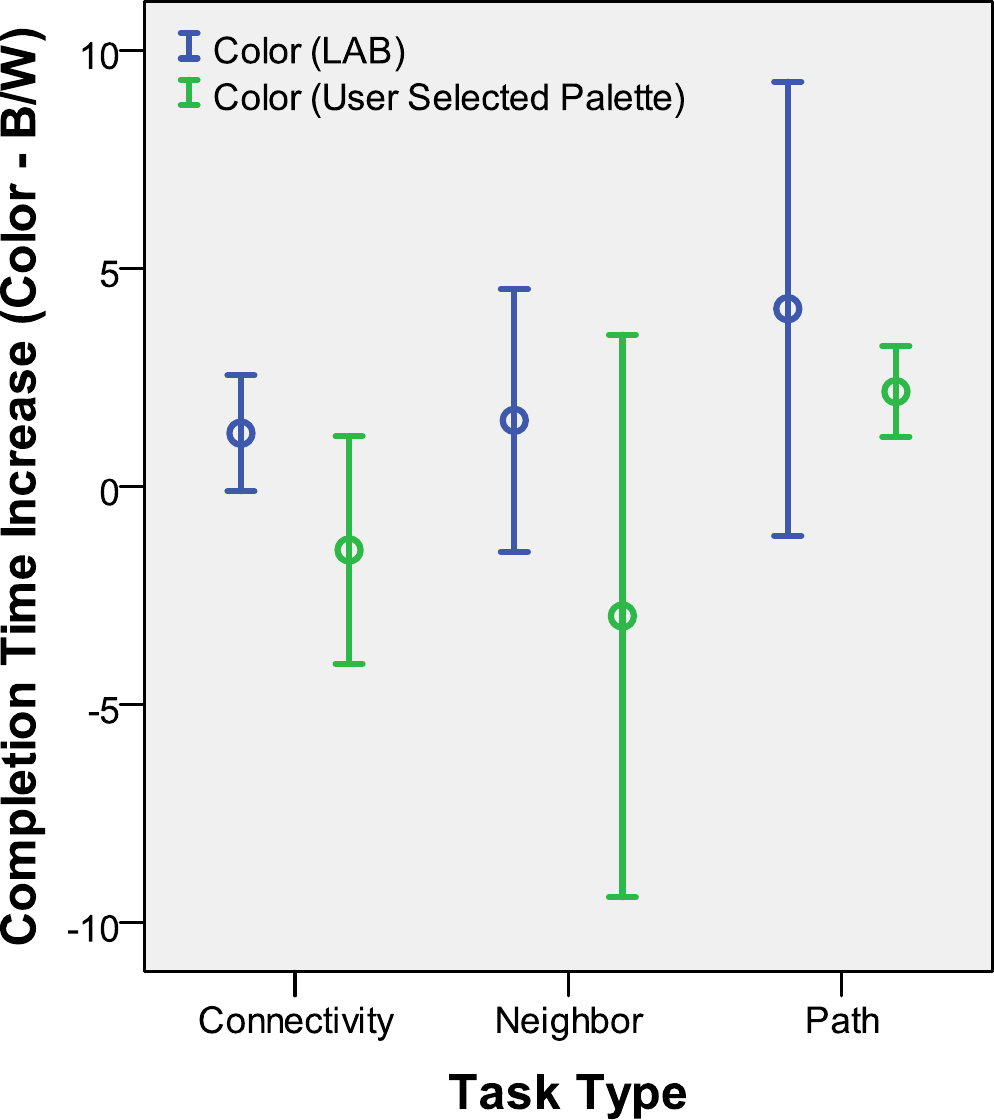}\\
(c)&(d)\\
\hspace{-.5cm}\includegraphics[width=0.25\textwidth]{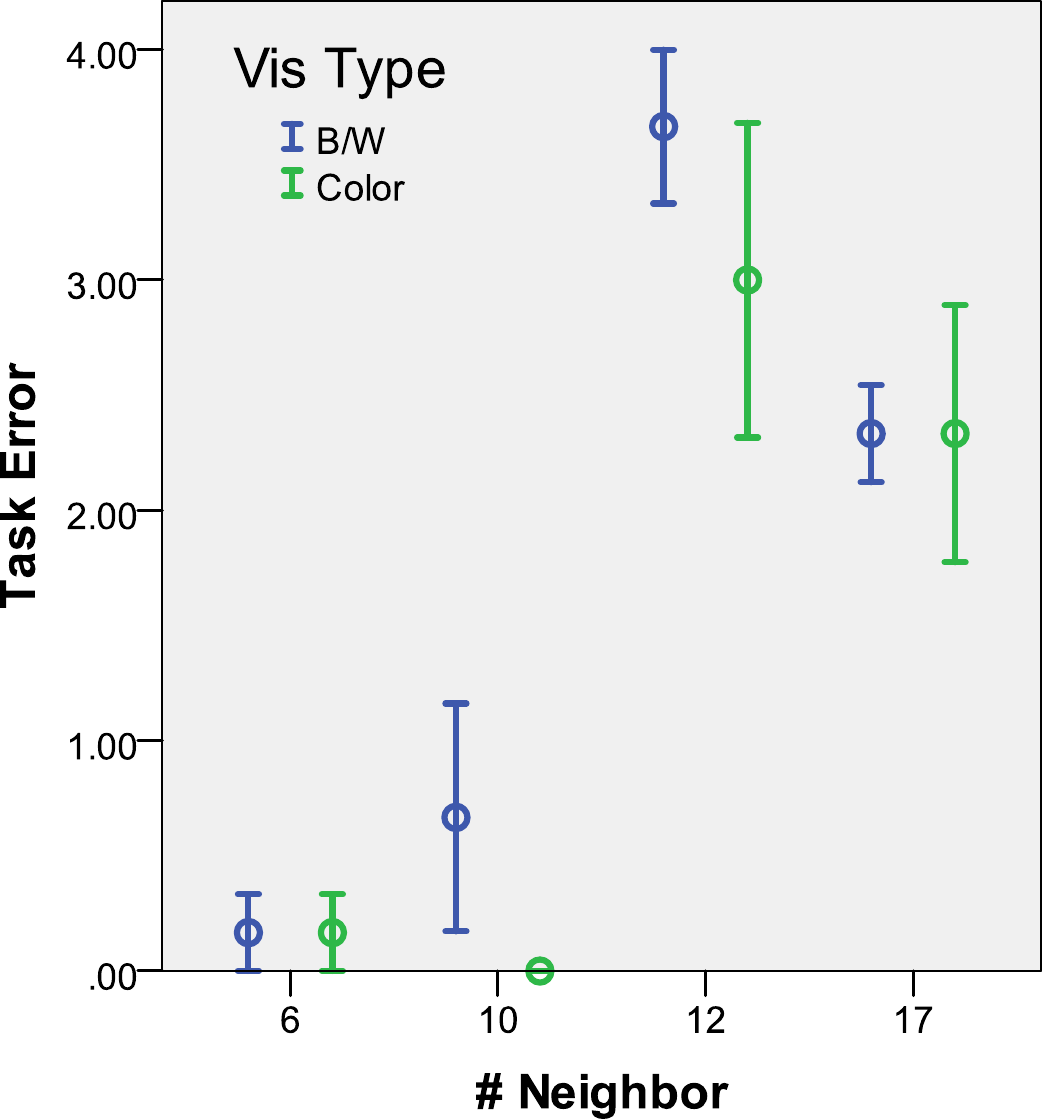}&
\hspace{-.25cm}\includegraphics[width=0.25\textwidth]{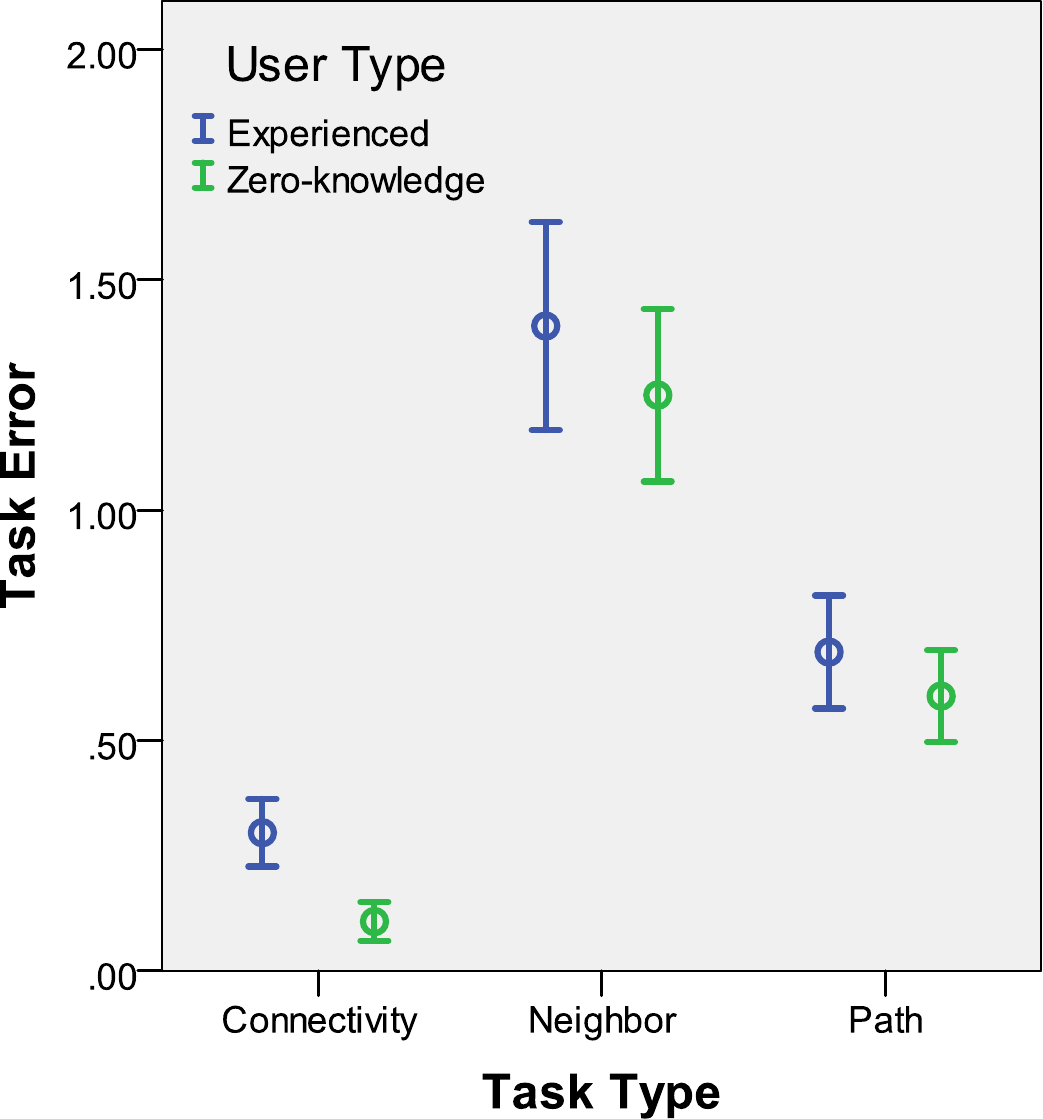}\\
(e)&(f)\\
\end{tabular}
\centering
\begin{tabular}{c}
\includegraphics[width=0.25\textwidth]{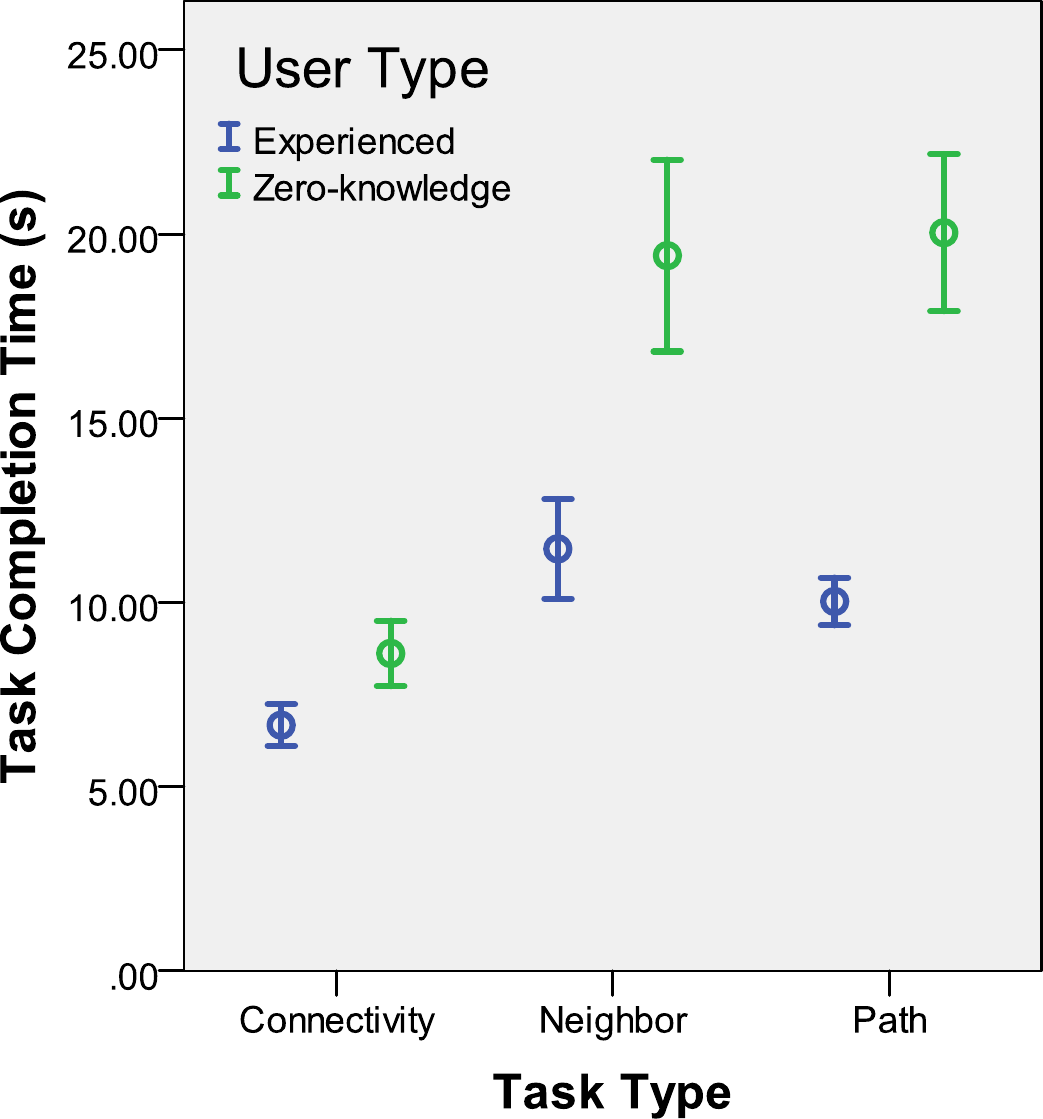}\\
~~~~~~(g)\\
\end{tabular}
\vspace{-.2cm}
\caption{Experiment results shown as error bars (Mean $\pm$ 1 Standard
  Error) on three task types separately: (a) Task error in B/W and
  Color groups; (b) Task completion time in B/W and Color groups; (c)
  Task error increase of the Color approach in LAB color space and
  with user selected color palette; (d) Task completion time increase
  in the same groups with (c); (e) Task error in finding neighbors by
  the number of actual neighbors; (f) Task error by user type, experienced v.s. zero-knowledge on node-link graphs; (g) Task completion time by user type.}
\vspace{-1cm}
\label{fig:UserStudyResult}
\end{figure}

We conducted a controlled experiment to study the effect of edge coloring on user's performance in fundamental graph-related tasks, such as visually following edges, finding neighbors and calculating the shortest path. Generally we compared two approaches, defined as two visualization types: the baseline graph drawing in black-white (B/W) and the improved graph drawing with edges colored by our algorithm (Color).

\textbf{Experiment design.} We recruited 12 participants (8 male, 4
female) for this paper-and-pencil experiment. 10 of the participants
were graduate students majoring computer science and the other 2 of
them were department assistants with no technology background. A half
of the participants ever had experiences on node-link graphs, one
student was even an expert on graph. The other half did not have
previous knowledge with the node-link graph. The experiment followed a
within-subject design that every participant entered all tasks with
both visualization types. To eliminate the learning effect over the
same task, we used two different layouts of the same graph data. The
design ended up full factorial on the choice of two visualization
types and two graph layouts. Each participant entered the same task
four times in total. The experiment order was randomized across
participants. Half of them completed the tasks first with the B/W
approach and then with the Color approach. Another half adopted the
opposite order. Further, in half of the time when participants were
given the colored drawing, the algorithm is fixed to use the LAB
palette. In another half, the participants selected their favorite
palette and completed tasks with the colored drawings generated by this palette.

Before participants took the experiment, a training session was held to make sure they understood each task and got familiar with both drawing approaches. The training session included one task from each task type on a simple graph. The organizer checked the answer of each training task and explained any ambiguity immediately. In the formal study session, we recorded participant's answer and completion time in each task. We did not distinguish the task reading time from the completion time, because all tasks were very short.

\textbf{Data and task.} Two layouts of the Zachary's Karate Club Graph were used. One was exactly the layout in Fig.~\ref{pipeline}. Another was rotated and re-labeled. Three types of graph-related tasks were designed:

\emph{T1 (Connectivity): Determine whether two particular nodes are connected by a direct edge;}

\emph{T2 (Neighbor): Estimate the number of nodes a particular node connects directly;}

\emph{T3 (Path): Estimate the minimum number of hops from a particular node to another, including the source and destination.}

On each type, four tasks were selected on each graph layout with
similar difficulty levels. To eliminate user's visual node querying time from their task completion time, we annotated the related nodes in each task on the corresponding graph layout before participants took the task.

\textbf{Result.} Results were analyzed separately on each task type. Significant level was set at 0.05 throughout the analysis.

\emph{Task error:} We computed the task error measure by the absolute deviation of the user's answer from the ground truth. \footnote{We also computed the task error rate (task error divided by the correct answer). However, there is little difference from the absolute error measure.} On Connectivity tasks, task error will be 1 if the answer is incorrect, 0 otherwise. After that, we applied the two-way repeated measure analysis of variance (ANOVA) test where the task error was the dependent variable, the visualization type and the choice of graph layout were two independent variables. On Connectivity tasks, the task error difference between B/W and Color group is statistically significant ($p < .01$). A close look at the error bar in Fig.~\ref{fig:UserStudyResult}(a) shows that the average task error of the Color group ($M=0.083$, $SE=0.04$) is less than 30\% of the B/W group ($M=0.292$, $SE=0.066$). On Neighbor tasks, the average task error of the Color group ($M=1.208$, $SE=0.204$) has a 15\% reduction from the average task error in the B/W group ($M=1.417$, $SE=0.204$), though statistically the difference is not significant. On Path tasks, the error performance of the two groups are almost the same.

\emph{Task completion time:} We applied a similar two-way repeated measure ANOVA test on task completion time. On all three types of tasks, there is no significant difference between B/W and Color groups. The detailed task completion time distribution is shown in Fig.~\ref{fig:UserStudyResult}(b).

\emph{Effect of color palette:} On each participant, we computed the increase of task error and completion time of the Color approach over the B/W approach on the same task and graph layout. We compared these measures between two groups: one applying the fixed color palette (LAB) to generate the drawing and another applying the user selected color palette from six candidates. Results are shown in Fig.~\ref{fig:UserStudyResult}(c) and Fig.~\ref{fig:UserStudyResult}(d). Though in all task types there is no significant difference between the two groups, we observe that on Connectivity and Neighbor tasks, using user selected palette leads to much smaller increases in both task error and completion time, which corresponds to solid performance improvement.

\textbf{Analysis.} The user study results demonstrate that the edge coloring technique on graph drawings can improve user's performance in identifying 1-hop graph connectivity significantly. This is also echoed by the subjective feedbacks from our participants. Most of them found the colored graph much clearer in showing the graph connectivity. While in the B/W drawings, they found it hard to distinguish the edges that are crossed by or closed to other edges at small angles. On the third task to quantify the shortest path, the Color approach does not have a comparative advantage to the B/W approach. This is reasonable because most cases in this task involve a rather long path. It is difficult for people to figure out the exact path length only by eye, even if the graph is colored. Lastly, the intermediate task to estimate the number of 1-hop neighbors returns some surprising results. We have expected the Color approach to be significantly better than the B/W approach. However, only a 15\% reduction in the task error is observed in average and the difference is not significant. To account for this result, we drilled down to the detailed cases and plotted Fig.~\ref{fig:UserStudyResult}(e) to show the relationship between the Neighbor task error and the number of actual neighbors. It is clear that the task error is large on nodes with a higher number of neighbors, and smaller on other less-connected nodes. On the nodes with a medium number of neighbors (10, 12), the Color approach is better than the B/W approach; while on the node with the highest number of neighbors (17), there is no error reduction for the Color approach. We also asked the participants making the most errors about the challenges in completing Neighbor tasks. Quite a few of them mentioned the same reason: most crossing angles between the edges on the target node are too small, so that they can not determine the exact number of edges on that node, no matter the graph is colored or B/W.

Possible concerns on the user study design are the small number of
participants enrolled, and whether the diversified user background can
interfere with our main result. We argue that our experiment has a
fully within-subject design. Each participant is tested and measured
24 times, adding up to 288 entries in both participant's answer and
completion time. This is sufficient to get an initial idea of the edge
coloring effect. We also looked at the impact of the user background.
In Fig.~\ref{fig:UserStudyResult}(f), it is shown that participants
with zero knowledge on the node-link graph actually made fewer errors
than other experienced users. Further investigation on their task
completion time narrates the potential reason: the zero-knowledged
participants spent much longer time than the experienced users
(Fig.~\ref{fig:UserStudyResult}(g)) -- they
are simply more careful in taking the tasks. By a two-way ANOVA test,
we found that there is little interaction between the user background
and the visualization type on the task error performance. The
advantage of edge coloring applies unbiasedly to both experienced and
first-time users. Comparing Fig.~\ref{fig:UserStudyResult}(a) and
\ref{fig:UserStudyResult}(f), it can be observed that the task error
improvement with the edge coloring is similar in magnitude 
to that brought by more carefully taking the task and spending more time on the questions.

%% if specified like this the section will be ommitted in review mode
%\acknowledgements{
%The authors wish to thank Frank Van Ham for his valuable comments during the time of this work.}

\section{Discussions\label{sec_limit}}

The approach of coloring edges for disambiguating drawings has its limitations. Our working assumption is that 
the drawing is to be display as a static image on paper, or on screen. In case when an interactive environment
is available, interactive techniques such as ``link sliding'' and ``bring \& go'' \cite{Moscovich_2009_bring_n_go}
could be more effective. In such a situation,
the algorithms proposed here can be used as an additional visual aid to the interactive techniques.

While the algorithm proposed here can run on relatively large graphs, our experience is that for graphs
with a lot of edges, a static image is insufficient to allow the user
to clearly see and follow each edge.
Therefore our approach is best suited for small- to medium- sized graphs. Typical usage scenarios are
illustrations of diagrams, such as computer or biological networks.

Finally, we note that sometimes edge colors are used to encode
attributes on the edges. To apply our approach without interfering with
the need to display such attributes, edges can be differentiated using
dashed lines of different style and/or thickness, using the same
algorithm in this paper. This can be achieved by mapping different
line styles to 1D or 2D spaces.

\section{Conclusions\label{sec_conc}}

Edge crossings, particularly those at small crossing angles, are known to be detrimental to the 
visual understanding of graph drawings.
This paper proposes an edge coloring algorithm for disambiguating edges that are in collision because of
small crossing angles or partial overlaps. The algorithm,  based on a branch-and-bound procedure applied to a space
decomposition of the color gamut, generates color assignments that maximize color differences of
the colliding edges, and works for both continuous color space and
discrete color palettes. 
The algorithm can also be applied to generate coloring for disambiguating virtual maps.
Our user study found that
that  coloring edges in graph drawings helped user's performance in 1-hop graph connectivity task significantly.
Consequently we have made the CLARIFY code available as part of an open source software.

For future work, we plan to investigate better initial coloring strategies,
before applying the CLARIFY algorithm. These include
coloring high degree nodes first, or use a strategy similar
to the register allocation algorithm~\cite{George_1996_register_assignment}.

\bibliographystyle{abbrv}
%\bibliography{paper}

\end{document}